\documentclass[useAMS,usenatbib]{mn2e}
\usepackage{psfig, subfig, graphicx}
\usepackage[usenames,dvipsnames]{ xcolor}
\usepackage{multirow}
\usepackage{Times}
\usepackage[utf8]{inputenc}

\title[Finding Exoplanets Orbiting Young Active Stars. I. Technique]{Finding Exoplanets Orbiting Young Active Stars. I. Technique}

\author[V. E. Moulds, C. A. Watson, Xavier Bonfils, S. P. Littlefair, Elaine Simpson]{V. E. Moulds$^{1}$\thanks{E-mail:
vmoulds01@qub.ac.uk}, C. A.
Watson$^{1}$, X. Bonfils$^{2}$, S. P. Littlefair$^{3}$ and E. K. Simpson$^{1}$\\
$^{1}$Department of Physics and Astronomy, Queens University Belfast, Belfast, BT7 1NN, Northern Ireland\\
$^{2}$UJF-Grenoble 1/CNRS-INSU, Institut de Plan\'{e}tologie et d'Astrophysique de Grenoble (IPAG)UMR 5274,38041 Grenoble, France\\
$^{3}$Department of Physics and Astronomy, University of Sheffield, Sheffield S3 7RH}
\begin{document}

\date{\center{\Large Submitted for publication in the Monthly Notices
    of the Royal Astronomical Society \\
\vspace{.5cm} \today}}

\pagerange{\pageref{firstpage}--\pageref{lastpage}} \pubyear{2012}

\maketitle

\label{firstpage}

\begin{abstract}
Stellar activity, such as starspots, can induce radial velocity (RV) variations that can mask or even mimic the RV signature of orbiting exoplanets.  For this reason RV exoplanet surveys have been unsuccessful when searching for planets around young, active stars and are therefore failing to explore an important regime which can help to reveal how planets form and migrate.  This paper describes a new technique to remove spot signatures from the stellar line-profiles of moderately rotating, active stars (v$\sin$i ranging from 10 to 50 kms$^{-1}$).  By doing so it allows planetary RV signals to be uncovered.  We used simulated models of a G5V type star with differing dark spots on its surface along with archive data of the known active star HD49933 to validate our method.  The results showed that starspots could be effectively cleaned from the line-profiles so that the stellar RV jitter was reduced by more than 80\%.  Applying this procedure to the same models and HD49933 data, but with fake planets injected, enabled the effective removal of starspots so that Jupiter mass planets on short orbital periods were successfully recovered.  These results show that this approach can be useful in the search for hot-Jupiter planets that orbit around young, active stars with a v$\sin$i of $\sim$ 10 - 50 kms$^{-1}$.
\end{abstract}


\section{Introduction}

Currently over 800 exoplanets have now been found through a variety of techniques e.g. via transits, radial velocity (RV) variations and microlensing.  The most fruitful of these is the RV method which has been responsible for discovering approximately 70\% of exoplanets to date.  High-precision spectrographs with an accuracy of a few ms$^{-1}$ can easily detect the hundreds ms$^{-1}$ amplitude of the RV signature of short period hot-Jupiter planets.  At this level of accuracy, the amplitude of the signal can only be hidden by important stellar noise.\\  \cite{Desort:2007di} showed that stellar noise, such as starspots, can produce RV shifts that can mimic or mask out planet signals.  The limiting parameter for finding planets in the presence of such activity is the ratio between the amplitude of the planetary and activity signal.  This means that stellar activity on stars similar to the Sun, which exhibit an RV jitter of several ms$^{-1}$, will only affect the detection of low mass, long orbital period planets (e.g. \citealt{Lagrange:2010di}).  Whereas, for young, active stars were stellar jitter can be of the order of kms$^{-1}$, the search for planets is limited to high mass planets and in particular those on short orbital periods (e.g.\citealt{Paulson:2004di}, \citealt{Paulson:2006di}).\\There have been several cases where planets have been announced and then subsequently retracted after starspots were discovered to be the source of the RV signature.  Notable examples of this are TW Hydrae (\citealt{Setiawan:2008di}, \citealt{Huelamo:2008di}), BD+20 1790 (\citealt{Hernan:2010di}, \citealt{Figueira:2010di}) and HD166435 (\citealt{Queloz:2001di}).  Due to these problems, RV surveys often exclude active stars.  Young stars with convective outer envelopes tend to be faster rotating than older stars and therefore more active.  According to the data available on http://exoplanet.eu/catalog/ this has meant that only 57 planets with ages less than 1 Gyr are known to date.  \\Targeting young planets is important for understanding the formation and evolution of planetary systems.  Since the core accretion model (\citealt{Pollack:1996di}) predicts longer formation timescales than the disk instability model (\citealt{Boss:1997di}) then a detection of a young planet would provide information about the formation mechanism itself.  Several people have already conducted surveys for young planets (e.g. \citealt{Paulson:2004di}, \citealt{Huerta:2008di}) but without any success due to small sample sizes as well as the high stellar noise problem.   \\Searching for companions around young stars is also important for understanding the lack of substellar objects with masses greater than 20 $M_J$ and in orbits closer than 3 AU, i.e. the brown dwarf desert.   \cite{Grether:2006di} found that 16\% of nearby sun-like stars had companions with orbital periods less than 5 years and of these less than 1\% were brown dwarfs.  It has been suggested by \cite{Armitage:2002di} that the brown dwarf desert could be the result of orbital migration.  Since the timescale for catastrophic migration is of the order of 1 Myr, they predict that young stars would not have destroyed their brown dwarf companions and have an order of magnitude more brown dwarf companions than main-sequence stars.  A survey for substellar companions around young stars could help with this theory and our understanding of the brown dwarf desert.   \\Finding a solution to the stellar activity problem is not only important for conducting young planet RV surveys but also for transiting missions.  The Kepler space mission was launched in 2009 and monitors over 150000 stars with early data showing numerous transiting planets to be of low mass (\citealt{Borucki:2011di}).  With future planned missions such as TESS (\citealt{Ricker:2010di}), which will monitor over 2 million stars in its lifetime, the number of exoplanets, especially small ones, will continue to increase.  However transiting surveys, unlike RV surveys, do not initially exclude active stars from their missions.  In addition, early Kepler results show that approximately 50\% of the stars that Kepler observed during the first month of the mission are more active than the Sun (\citealt{Basri:2010di}).  This could result in difficulties in the RV follow-up work to confirm their planetary nature particularly around more active stars.  \\Indeed, this has already been found to be the case for the Super-Earth planet transiting the active star CoRoT-7 in 2009.  The activity of the star completely masks out the planetary signatures which has led to problems when analysing the RV data with separate groups finding different planetary solutions.  \cite{Queloz:2009di} found 2 planetary signals giving masses of 4.8~$M_{Earth}$ and 8.4~$M_{Earth}$ for the planets, whereas \cite{Hatzes:2010di}, analysis of the same data, suggests the star has 3 planets with masses of  6.9~$M_{Earth}$,  12.4~$M_{Earth}$ and 16.7~$M_{Earth}$.  \\Over the last couple of years a lot of effort has been put into finding a solution to this problem.  One solution is to use the relationship between bisector-span and RVs in order to remove activity signatures from the data, e.g. \cite{Boisse:2009di} used this technique on HD189733 to improve the RV jitter from 9.1~ms$^{-1}$ to 3.7~ms$^{-1}$.  However, this cannot be applied to all systems.  \cite{Desort:2007di} showed that this correlation breaks down for stars with $v$sin$i$ that is lower than the resolution of the spectrograph, due to the fact the spot cannot be resolved easily.  There are also problems when stars have multiple spots, due to spot distortions compensating for each other.  If other effects contribute more significantly to the RV variations (such as a planet signal) then the correlation may in fact be absent.  \\Other methods, such as pre-whitening techniques (\citealt{Queloz:2009di}), Fourier Analysis (\citealt{Hatzes:2010di}) and Harmonic decomposition (\citealt{Boisse:2011di}) have also been used to uncover planet signals from stellar jitter.  These methods are all based on analysing the RV data in order to identify spurious RV signals due to activity and remove them.  There are, however, some limitations when using these techniques.  They require accurate knowledge of the stellar rotation period and in the case when the planetary orbit is close to the stellar rotation period then separating out the signal can be extremely difficult.  Generally a long time series of RV points is required in order to measure the rotation period and so considerable observational time is important.  However, when all these criteria are met, harmonic decomposition has been found to pull out planetary signals that are a third smaller in amplitude than the activity signal (\citealt{Boisse:2011di}).  \\We have developed a technique to compliment these existing methods.  Instead of removing the activity signal from the RV points we remove the spot signatures from the actual lineprofiles directly.  This method does not require knowledge of the stellar rotation period or a long time series of data, provided the  planetary timescale is short and the jitter/planetary signal ratio is small.  This is the first paper of a series, in which we shall outline this novel spot removal technique.  A forthcoming paper will apply this technique to FIES observations of a sample of young, moderately-fast, rotating stars.\\Section 2 of this paper provides a detailed description of this spot removal technique.  We then tested this technique on model stars with varying spots and planets.  In section 3 we describe the construction of these models, and the results of our simulations are presented in section 4.  In section 5 we apply our spot removal technique to archival data of the known active star HD49933.  A discussion of this technique in comparison to other spot removal methods is provided in section 6, with future work detailed in section 7.  Finally, we summarise our findings in section 8.

\section[]{Spot Removal Technique}

Our technique is based on the fact that a planet and spot have different effects on the line-profile.  A planet causes the line-profile to shift in wavelength or velocity whereas a spot distorts the actual shape of the line-profile resulting in an \textit{apparent} RV shift.  Spot bumps in the lines can be resolved when rotational broadening dominates the shape of the line profile, typically for $v$sin$i$'s greater than the resolution of the spectrograph.  We have written a code, Clearing Activity Signals In Line-profiles (ClearASIL), to assess the stellar absorption line-profiles and remove any spot features, effectively cleaning the RVs to allow any planet signatures to be uncovered.  ClearASIL is based upon the method used by \cite{Collier:2002di} on AB Doradus, to directly track starspots in order to assess stellar differential rotation.   

\subsection{Least Squares Deconvolution (LSD)}
In order to resolve the typically small spot bumps, an absorption line with a signal-to-noise ratio (SNR) of several hundred is required.  Although large format CCDs and modern \'{e}chelle spectrographs enable high resolution spectra to be obtained spanning a large wavelength range (typically 3500 to 7000 \AA\, for optical spectrographs such as FIES) the SNR of a single line is often not adequate.  This is due to the requirement of short exposure times in order to reduce the effects of blurring of the spot features, and is further compounded for faint targets.\\  Each of the 1000's of photospheric lines contained in an \'{e}chelle spectrum are assumed to be affected in a similar way by the presence of spot features in the stellar photosphere.   This assumption enables the application of multiline techniques to extract common physical information from many spectral lines and determine an average profile with a high SNR.  LSD is the method used in this paper to compute a high SNR line-profile from the thousands of spectral lines available in a single \'{e}chelle spectrum.  It was first implemented for use on polarimetric Stokes V spectra by  \cite{Donati:1997di} and has since been successfully employed in other areas of astronomy, such as Doppler imaging.\\  However, the simplifying assumption that all spectral lines have the same shape, i.e. are independent of wavelength, is not valid for spectral lines formed at different heights which will be impacted by spots differently and also for bluer line-profiles were spot features will appear larger due to the increased temperature contrast.  Therefore, the LSD technique is only applied after spectral regions containing chromospherically sensitive and strong spectral lines were removed from the data in this paper. \\  In this paper the LSD code used was written by C.A.Watson (\citealt{Watson:2006di}).  Using a stellar line list from the Vienna Atomic Line Database (VALD) and the method of least squares, the average profile is the one that gives the optimal fit to the spectra when convolved with the line list.  The resulting line-profile has a SNR that is typically 20 to 30 times higher than a single isolated line.

\subsection{Radial Velocity Calculation} 
\label{Sec:RVCalc}
The LSD line-profiles can then be used to measure the RVs for the data, and this is achieved by fitting a gaussian to the LSD line-profiles and taking the peak of this gaussian.  This technique for calculating the RVs corresponds to the widely used cross-correlation method whereby a template spectrum is cross correlated with the target spectrum and the peak of a gaussian fitted to this cross-correlation function is measured. \\
The LSD line-profile can be approximated by a gaussian of the form, 

\begin{equation}
\label{eq:gauss}
p_{i} \, = \, A \, \exp \, \left( \, \frac{(\, v_i \, - \, \mu \,)^{2}}{2\sigma^{2}} \, - \, y_{off} \right)
\end{equation}

where $p_{i}$ is the model flux value at the velocity value $v_i$.  The four parameters which define the gaussian are the area under the curve, $A$, the variance (i.e. the width of the line), $\sigma$, the mean (i.e. the peak position), $\mu$, and the continuum flux level, $y_{off}$.   \\A least squares minimization technique is used to determine the model parameter values that give the best fit to the data.  The measure of the goodness of fit between the model and observed data is given by the following $\chi^{2}$ statistic,

\begin{equation}
\label{eq:chi-squared}
\chi^{2} \, = \, \sum_{i} \, \left( \, \frac{f_{i} \, - \, p_{i}}{\sigma_{i}} \right)^{2}
\end{equation}

where $f_{i}$ and $p_{i}$ correspond to the observational and model flux at the $i^{th}$ data point and $\sigma_{i}$ corresponds to the observational error for the $i^{th}$ data point. \\The best fit model is found by minimising the $\chi^{2}$ statistic using a non linear least squares technique, specifically the Levenberg-Marquardt algorithm.  This is an iterative process that rapidly converges on a numerical solution to the minimisation of the function.  As with all least-squares techniques, an initial guess of the parameters is required.  This must be relatively close to the true value in order to avoid the results returning a local minima.  The initial parameters are chosen by the user based upon the shape of the line-profile.  In this case the C version of the IDL MPFIT routine (\citealt{Markwardt:2009di}) was employed.\\This method for measuring the RVs using the LSD line-profiles was tested on our model data as detailed in Section~\ref{Sec:RVtest} and the results showed this method to be accurate. \\

\subsection{ClearASIL}
The high SNR mean line-profiles (in this case obtained using the LSD technique) are then input into ClearASIL. The code operates on the basis that RV variations of a time series of line-profiles are either due to spots only or due to both spots and an orbiting planet.   ClearASIL processes the line-profiles according to both these theories and then tests to see which assumption best removes the spot features.   This is a three stage decision process that is outlined in more detail below.   The code iterates over this process using the resulting average line-profile from the best spot removal method in the next iteration.  At the end of each iteration a goodness of fit test ($\chi^{2}$) between the resulting line-profiles from each fitting method and their average is calculated to determine the correct fitting technique.  When further iterations do not show any improvements to this goodness of fit test then ClearASIL stops.   

\subsection{Step 1 - Planet Fitting Technique}
This method is only employed when ClearASIL is testing the assumption that the data contains both a spot and planet.  If a planet is present then the line centers will be shifted relative to one another.  If we correctly remove this shift then the line centers should lie close to one another and the average profile of these shifted lines should have a similar width and central velocity peak compared to each of the individual lines.  On the other hand if we shift the lines by the wrong amount then their average will be orbitally smeared and hence appear broader when compared to the individual lines.  A goodness of fit test between the average line and each individual orbitally corrected line-profile is therefore an appropriate measure to find the best planet signal.\\
If the planet has zero orbital eccentricity then the RV variation of the star over time will be a smooth sinusoid.  The majority of hot-Jupiter planets have been found in circular orbits which is thought to be a result of short circularization time-scales for these close orbiting planets (\citealt{Jackson:2008di}).  As the work in this paper concentrates on detecting hot-Jupiter planets then a valid approach in our method is to assume zero orbital eccentricity.  With this assumption in mind the code generates sinewaves with various periods, amplitudes and phase offsets.  Shifting the line-profiles according to each sinewave and measuring the goodness of fit between these shifted lines and the average determines the best model planet for the data.  It is important to remove the planet signature at this stage of the fitting procedure so as to enable accurate removal of any spot features present in the subsequent steps.    

\subsection{Step 2 - Spot Fitting Technique}
\label{sec:spotremoval}
\begin{figure}
\includegraphics[trim = 10mm 10mm 10mm 50mm, clip,width=8.5cm, height=8.5cm]{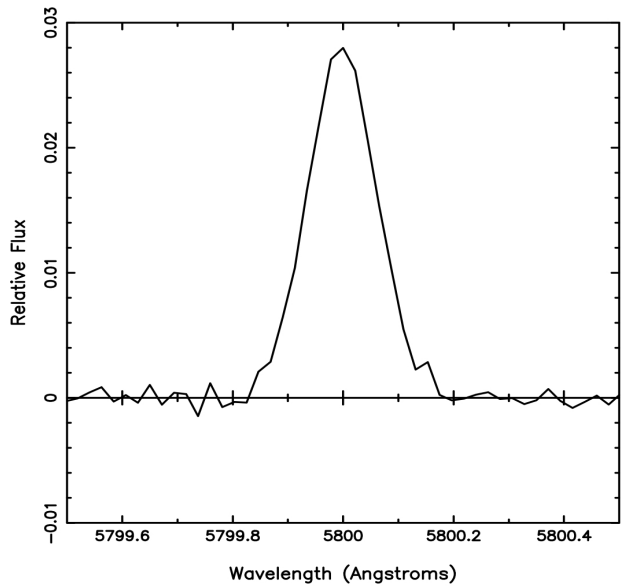}
\caption{This is the spot signature after the time-averaged line-profile has been subtracted from one of the individual spotty line-profiles generated by the model star shown in Fig.~\ref{fig:1Spot}.}
\label{fig:Resid}
\end{figure}

The observable signature of a dark spot on the surface of the star is a bright bump in the photospheric line-profile.  To isolate these bright bumps in the LSD profiles we employed the technique of line-profile subtraction. \\
Subtracting a line-profile with no spot features from each inidividual line-profile would result in residuals that contained noise with any spot bumps that were present superimposed, similar to Fig.~\ref{fig:Resid}.  Unfortunately it is impossible to obtain a completely immaculate line-profile from the data and so we generated a psuedo-immaculate profile by averaging all the line-profiles over the course of the observations.  Averaging the lines will cause any spot features present to be smeared out and diluted.  So although the average line-profile will not be completely free of spot features compared to the individual line-profiles spot features present will be severely diminished.  \\ 
This method utilises the approximation that spot features present in the residuals are Gaussian-like in shape as discussed by \cite{Collier:2002di}.  The residuals are scanned for any peaks above this noise level and these features are fitted with a Gaussian using the Levenberg-Marquardt technique.  The code also checks the width of these features to ensure spikes in noisy data are not wrongly treated as spot features.  This enabled the spot bumps to be isolated from the noise features.  These isolated spot bumps were then subtracted from the individual line-profiles to generate a cleaner data set.     

\subsection{Step 3 - The Optimal Fitting Procedure} 

Having fitted the line-profiles assuming that either spots and a planet are both present (i.e. fitting the lines using both steps 1 and 2 above) or that only spots are present (i.e. fitting the lines using only step 2 above) we now determine which of these techniques best models the data.  This is achieved by comparing the average of the cleaned line-profiles to the individual cleaned line-profiles for each method.\\
In the next iteration the average cleaned line-profile resulting from the best spot fitting procedure is used as a new psuedo-immaculate profile along with the original observed line-profiles to generate the residuals in step 2.  Hence with each iteration we are improving our `immaculate' profile enabling better spot identification and removal.  As more iterations are conducted the average line becomes cleaner and in our simulations (see Section~\ref{modelResults}) very closely resembles the `true' immaculate profile.  The RV of the line-profiles also becomes less noisy and, if any planet is present, it should be easily detected.  ClearASIL stops iterating when there is no longer any improvement to the average cleaned line-profile with subsequent iterations.  The RV of the resulting cleaned line-profiles at this iteration are then searched for any planet signatures. 
 
\section[]{Simulations}

To validate our spot removal method we generated a variety of model stars with differing spot and planet configurations.  This section gives a description of how these models were constructed.\\  

\subsection{Description Of Simulation Models}

\begin{figure}
\centering
\subfloat{\includegraphics[trim = 10mm 10mm 10mm 20mm, clip,width=3cm, height=4cm]{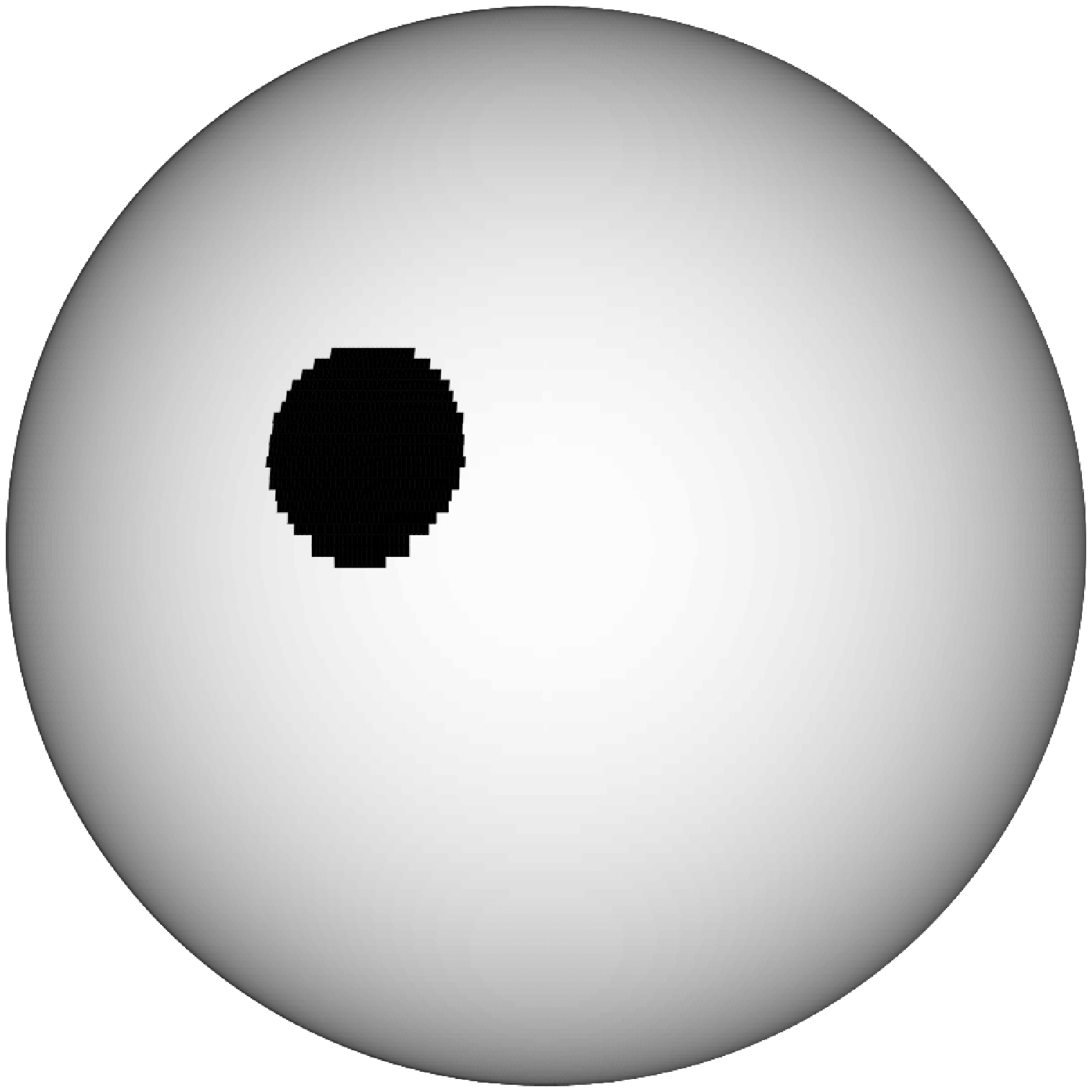}}
\subfloat{\includegraphics[trim = 10mm 10mm 10mm 20mm, clip,width=3cm, height=4cm]{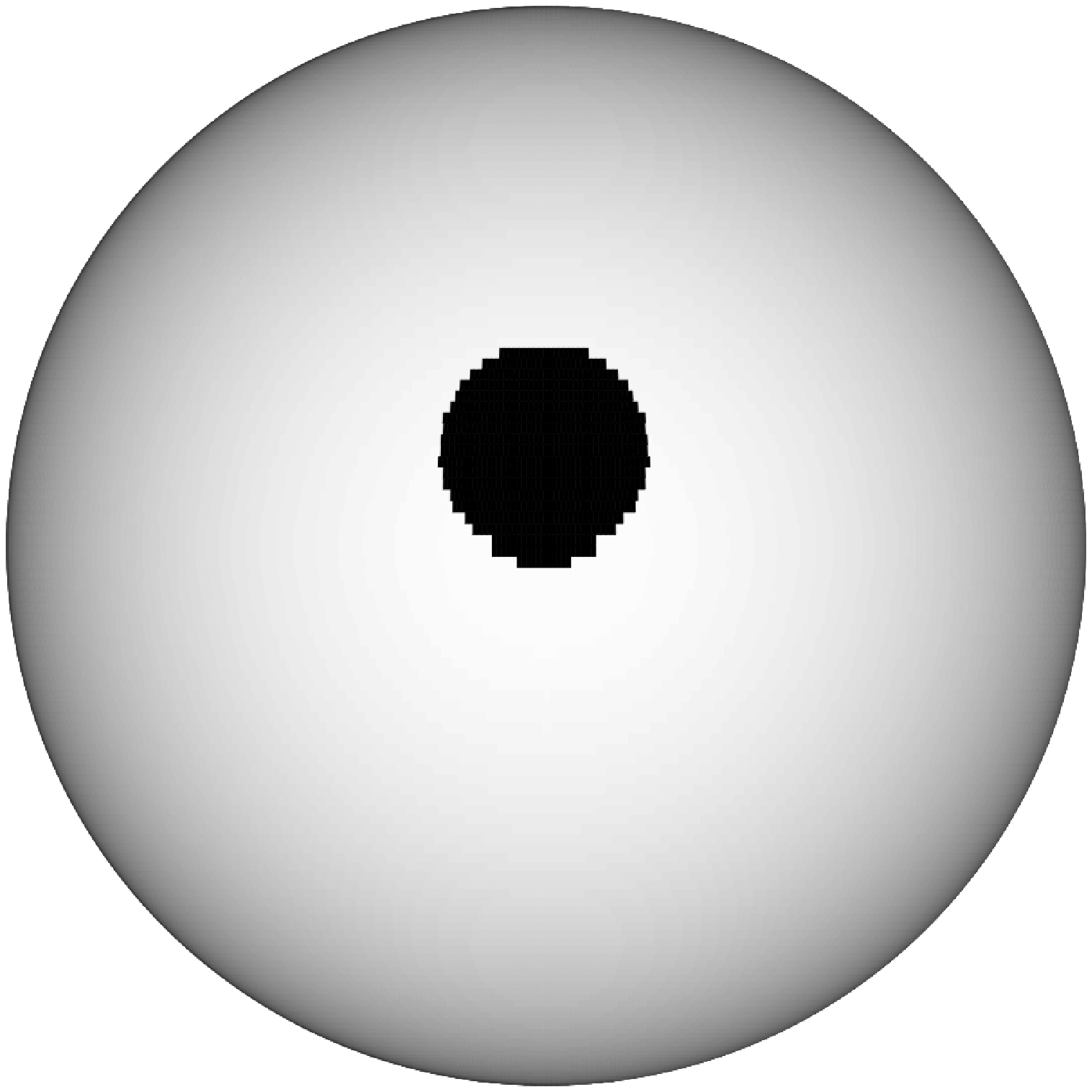}}
\subfloat{\includegraphics[trim = 10mm 10mm 10mm 20mm, clip,width=3cm, height=4cm]{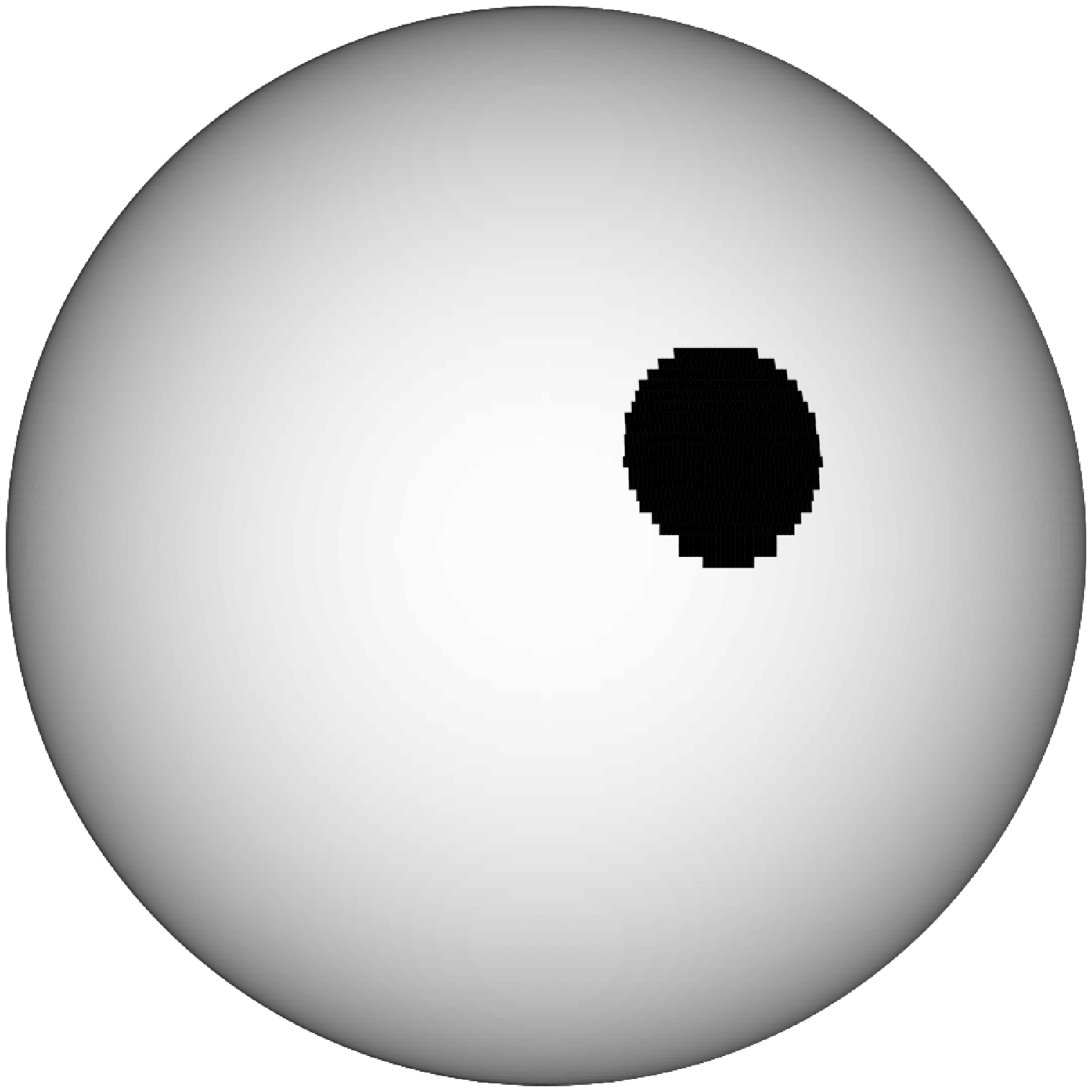}}
\\
\subfloat{\includegraphics[trim = 10mm 10mm 10mm 20mm, clip,width=3cm, height=4cm]{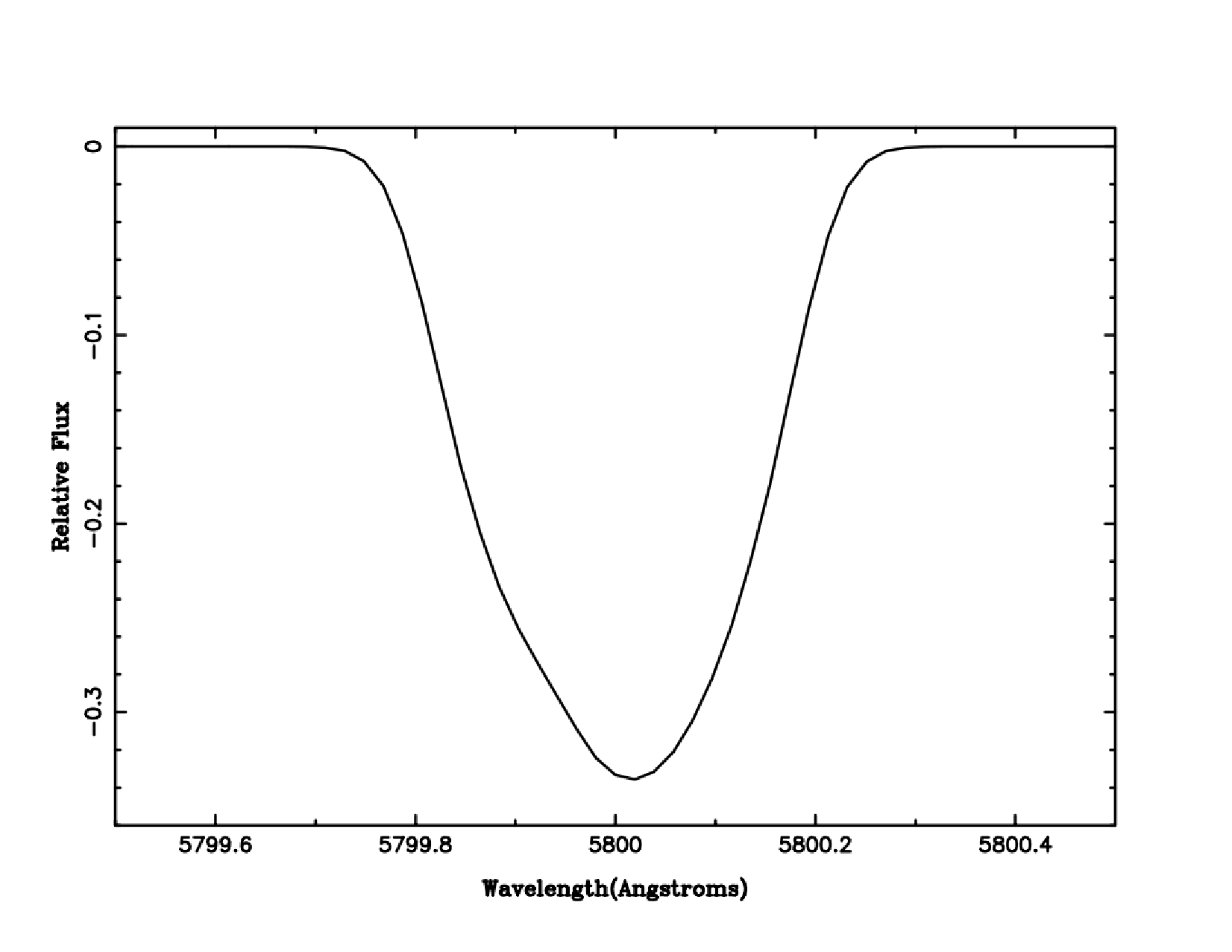}}
\subfloat{\includegraphics[trim = 10mm 10mm 10mm 20mm, clip,width=3cm, height=4cm]{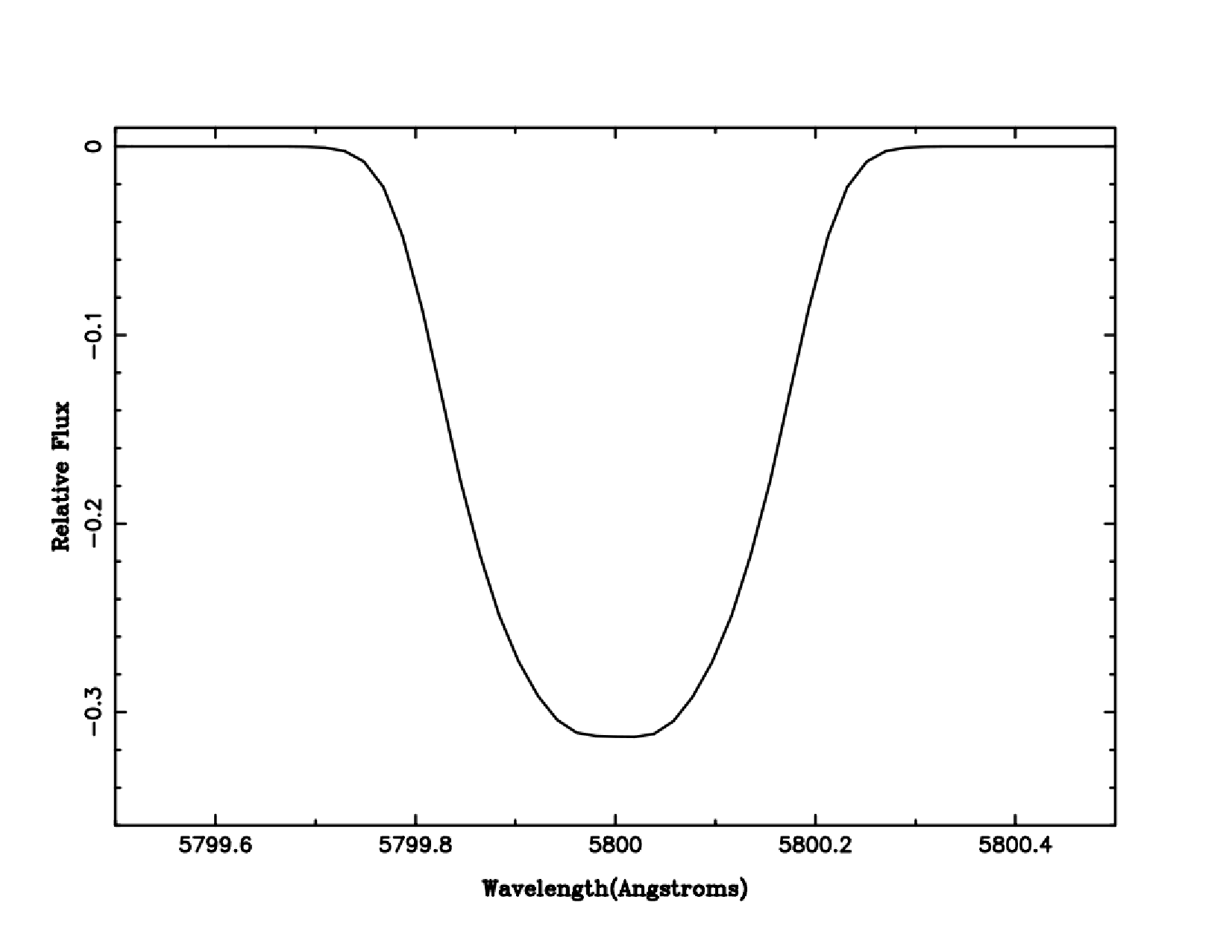}}
\subfloat{\includegraphics[trim = 10mm 10mm 10mm 20mm, clip,width=3cm, height=4cm]{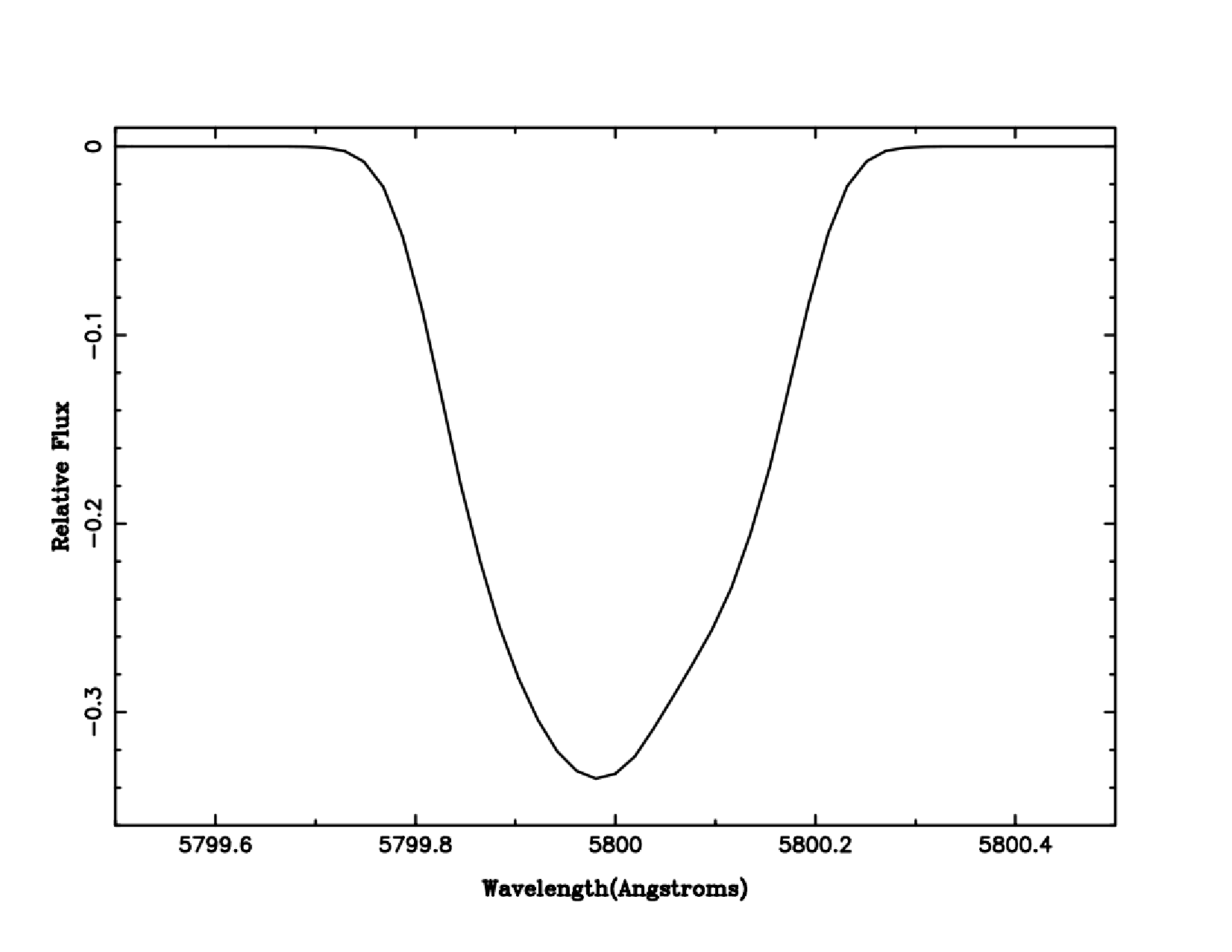}}
\caption[Rotationally Broadened Lines]{The effect of a rotating star spot on the rotationally broadened line-profile.  The top panels show the model star at different phases while the lower panel represents the line-profiles at this corresponding phase.}
\label{fig:RotatingSpot}
\end{figure}

We first generated a model star defined by 7 parameters: mass ($M_*$), radius ($R_*$), effective temperature ($T_{eff}$), rotational period ($P_{rot}$), inclination ($i$), microturbulent velocity  ($v_{micro}$) and macroturbulent velocity ($v_{macro}$).  We set the microturbulence to 1.5~kms$^{-1}$ for all the models used in this paper, while the macroturbulent velocity varies as a function of $T_{eff}$ and so has been calculated for each model using eqn 1. in \cite{Valenti:2005di}.  Including radial-tangential macroturbulence (as described in \citealt{Gray:2008di}) resulted in a more realistic model line-profile making it harder to resolve spot bumps.  Along with these parameters, the non-linear limb-darkening law of \cite{Claret:2000di} was used.  \\We then placed circular spots on the model star with free parameters given by the spot latitude ($\phi$) and longitude ($\theta$) on the stellar surface, the temperature of the spot ($T_{spot}$) and the spot size described by the fraction of the visible hemisphere covered by the spot ($f_r$).  We then added planetary reflex motion according to the planets mass ($M_{p}$) and orbital period ($P_{prot}$). We assume the hot-Jupiter planet has zero orbital eccentricity given that most hot-Jupiter planets have been found to have circular orbits.  \\We model the stellar surface as a series of quadrilateral tiles of approximately equal area.  Each tile is assigned a copy of the local (intrinsic) specific intensity profile, which includes micro and macroturbulence effects convolved with the instrumental resolution. A gaussian profile is used for the local (intrinsic) specific intensity profile and the equations in Chapter 17 of \cite{Gray:2008di} are used to model the micro and macroturbulence effects.  For the simulations in this paper we use an instrumental resolution of 4.6~kms$^{-1}$, which is equivalent to the FIbre-fed Echelle Spectrograph (FIES) on the Nordic Optical Telescope (NOT).  These profiles are scaled to take into account the projected area, limb-darkening, obscuration and the presence of a spot when relevant.  The intensity profile for each tile is then Doppler shifted according to the radial velocity of the surface element at a particular stellar rotation phase.  Summing up the contributions from each element gives the rotationally broadened profile at that particular stellar rotation phase.\\  Any tiles with a spot feature present will have a lower temperature than the surrounding tiles and hence a lower intensity.  We employ Planck's radiation law which gives the energy radiated by an ideal blackbody in a particular wavelength interval (the central wavelength of the model star line-profile is set to 5800\AA).  This law depends on the temperature of the blackbody and so if the tile has no spot feature then the energy radiated will be proportional to the effective temperature of the star.  Whereas if the tile has a spot then the intensity will be proportional to the spot temperature which is lower than the effective temperature of the star.  This results in an emission bump appearing in the rotationally broadened line-profile with a shape and position that is dependent on the temperature, size and position of the spot as shown in 
Fig.~\ref{fig:RotatingSpot}.\\

\begin{figure}
\includegraphics[trim = 10mm 10mm 10mm 20mm, clip,width=8.5cm, height=8.5cm]{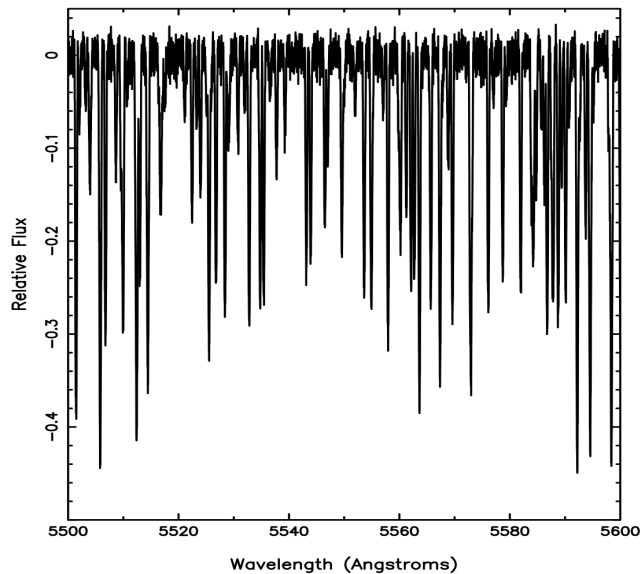}
\caption{Example portion of a synthetic spectrum generated by our model, assuming a G5V type star ($M_*$ = 1.05~$M_{sun}$, $R_*$ = 0.95~$R_{sun}$, $T_{eff}$ = 5657~K,  $P_{rot}$ = 3.2~days and $i$ = 51.8$^{\circ}$) with a 5\% spot having a peak SNR of 100 and a $v$sin$i$ = 15~kms$^{-1}$.}
\label{fig:Spectrum}
\end{figure}

We generate the resulting line-profiles for different epochs over the rotational phase.  Typically, we take 19 epochs to cover the rotational period.  If there is a planet orbiting the star then the rotationally broadened profiles are shifted in velocity according to the RV amplitude of the planet at that particular orbital phase.  The resulting line-profiles are then wavelength binned so that they are all on a common scale.  Each rotationally broadened profile is then convolved with an appropriate stellar line list, taken from VALD, with photon noise added in order to obtain a model spectrum as shown in Fig.~\ref{fig:Spectrum}.  The model spectrum covers the wavelength range, 4500 to 7000 \AA\, similar to the typical usable wavelength range of spectra taken with an optical spectrograph such as HARPS or FIES.\\  We have added photon noise assuming that the SNR is proportional to the square root of the flux.  Since our model spectra are normalised, in order to determine the flux as a function of wavelength we have performed a cubic spline fit to a synthetic spectrum.  The Pollux database provided the synthetic spectrum, of similar spectral type to our model star.  This enabled to correct relative photon noise to be calculated.  All the models described in this paper have been given a peak SNR of 100.\\

\subsection{Testing The Model}
\label{Sec:RVtest}

\begin{figure}
\includegraphics[width=7.5cm, height=7.5cm]{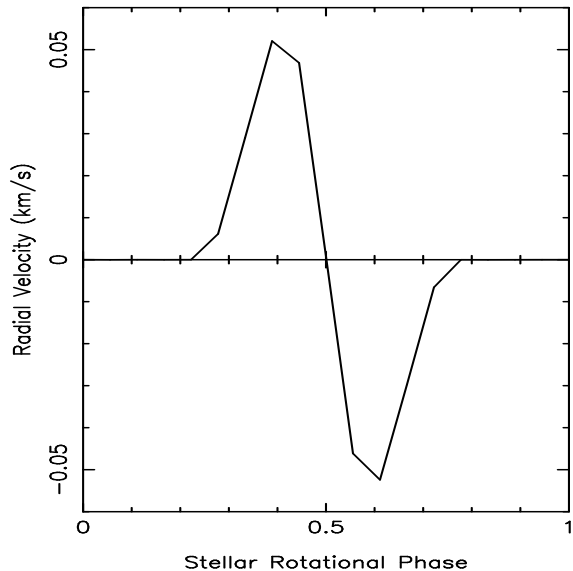}
\caption{The RV for a G2V type star seen face on with a dark spot ($T_{spot}$ = 0~K) covering 1.05\% of the visible surface at $\theta$ = 0$^{\circ}$.}
\label{fig:RVtest}
\end{figure}

We checked how accurate our models were by measuring the RVs for the line-profiles generated over one rotational period for a G2V model star and compared the resulting RV jitter to that expected using eqn. 1 in \cite{Saar:1997di}.  Again, we calculated the RVs by fitting a Gaussian to each individual line-profile and measuring the peak position (see Section~\ref{Sec:RVCalc}). Fig.~\ref{fig:RVtest} shows the RVs for a G2V model star ($T_{eff}$ = 5800~K) with $v$sin$i$ = 7~kms$^{-1}$ seen at an inclination of 90$^{\circ}$ with a spot placed at a latitude of 0$^{\circ}$, covering 1.05\% of the stellar surface and given a $T_{spot}$ of 0~K.  \citealt{Saar:1997di} predict a semi-amplitude of 47.54~ms$^{-1}$ for these parameters.  Our model produced an RV semi-amplitude of 52~ms$^{-1}$.  This gives us confidence in our model line-profiles and also in our method for calculating the RVs.\\ 

Satisfied with the method for generating our model line-profiles we computed a series of stellar models with varying spots and planets in order to validate ClearASIL.  The model star was given $M_*$ = 1.05~$M_{sun}$, $R_*$ = 0.95~$R_{sun}$, $T_{eff}$ = 5657~K,  $P_{rot}$ = 3.2~days and $i$ = 51.8$^{\circ}$.  The spot and planet configurations as well as the results for these models are discussed in the following sections.

\section[]{Model Results} 
\label{modelResults}
\subsection[]{No Spot Model}
\label{NoSpotModel}
\begin{figure}
\includegraphics[trim = 0.1mm 0.1mm 0.1mm 0.1mm, clip,width=7cm, height=7cm]{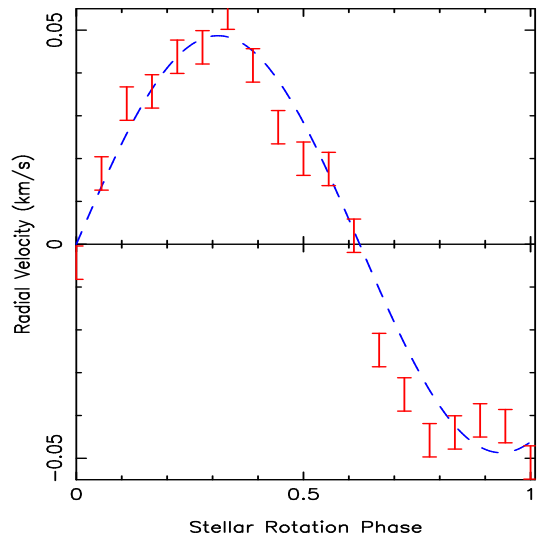}
\caption{The RVs for the model G5V type star with a 1$M_{j}$ mass planet present and no spots on the stellar surface.}
\label{fig:NoSpot}
\end{figure}

For the first test case a 1~$M_{J}$ planet on a 4~day circular orbit was placed around our model star.  No spots were present on the surface of the star and so the line-profiles should show a radial velocity shift solely due to the injected planet.  A total of 19 line-profiles covering one stellar rotation period (i.e. 3.2~days) were generated using the method described in section 3.  Note the error on the line-profiles corresponds to the SNR of 100 which is given to the model spectra.  This error is then propagated through the method for calculating the RVs and this is the case for all the model data described in this paper.  The RVs were found to have a semi-amplitude of 48.91$\pm$4~ms$^{-1}$ as shown in Fig.~\ref{fig:NoSpot}.  ClearASIL was applied to these line-profiles and it correctly found that no spot signatures were present.  Fitting the signal with a sinewave revealed a 1.0$\pm$0.04~$M_{J}$ planet on a 3.9$\pm$0.08~day orbital period.  This shows that the planet fitting method gives results that are consistent with the planet injected into this model.  \\ 

\subsection[]{1 Spot Model}
\label{1SpotModel}
\begin{figure}
\centering
\subfloat{\includegraphics[trim = 20mm 30mm 20mm 30mm, clip,width=5cm, height=6cm]{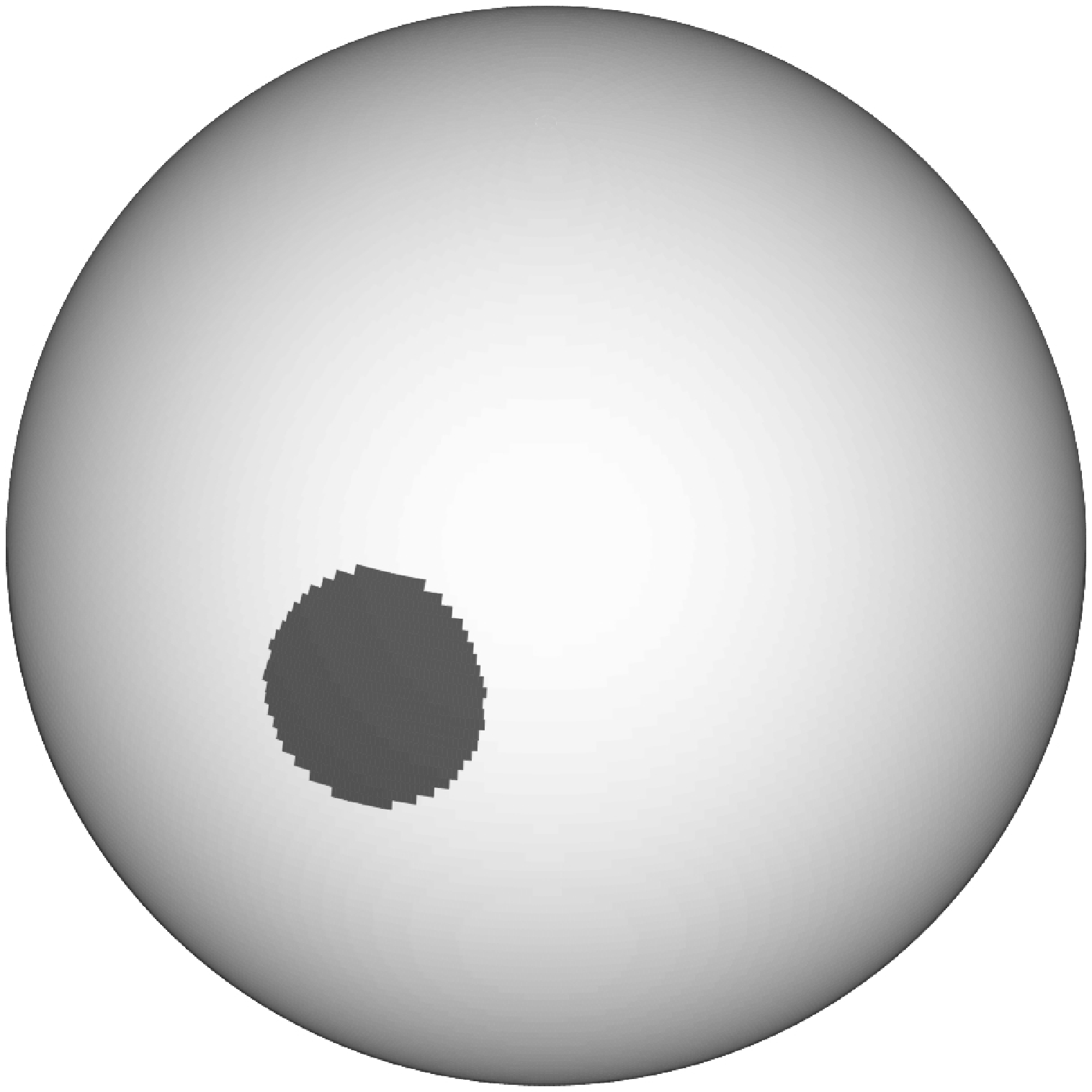}}
\\
\subfloat{\includegraphics[trim = 10mm 10mm 10mm 20mm, clip,width=4cm, height=6cm]{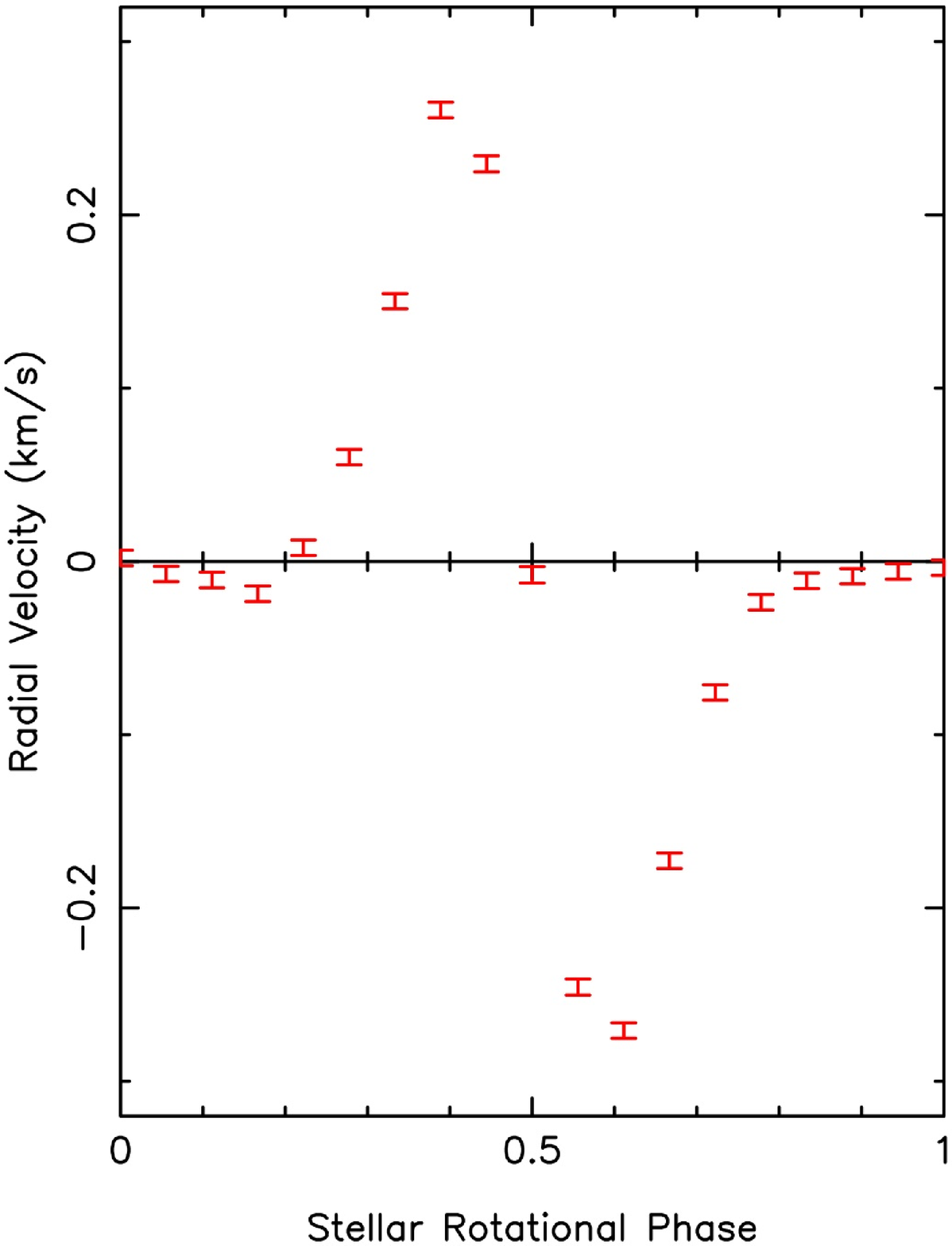}}
\subfloat{\includegraphics[trim = 10mm 10mm 10mm 20mm, clip,width=4cm, height=6cm]{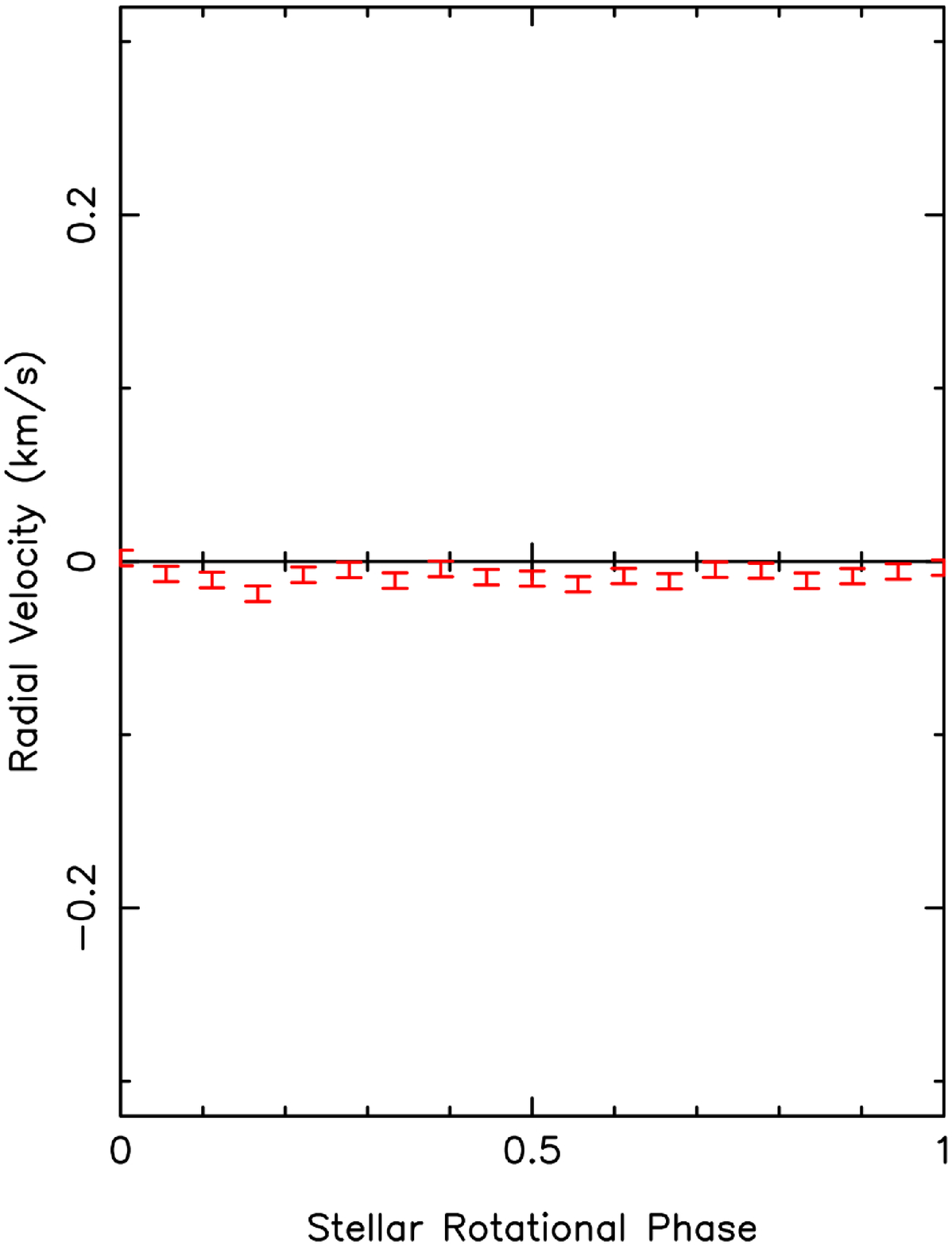}}
\caption{Top: The model G5V star with dark spot ($T_{spot}$ = 4600~K) covering 5\% of the visible surface at a latitude of 20$^{\circ}$.  Bottom Left Panel: The RV points determined using the 19 model line-profiles that were generated for this 1 spot star.  Bottom Right Panel: The RV results after using ClearASIL to clean spots from the model data.  The RV jitter due to the spot is reduced by 98\% in this test.}
\label{fig:1Spot}
\end{figure}

In the next test case we placed 1 spot on our model star at a latitude = 20$^{\circ}$ with $T_{spot}$ = 4600~K and covering 5\% of the visible stellar surface, as shown in the top of Fig.~\ref{fig:1Spot}.  A total of 19 line-profiles covering one stellar rotation period (i.e. 3.2~days) were generated using the method described in section 3.  The RVs were found to have a semi-amplitude of 265$\pm$5.7~ms$^{-1}$ as shown in the bottom left panel of Fig.~\ref{fig:1Spot}.  The amplitude of the RV signal is approximately the same as that caused by a 5~$M_{J}$ planet orbiting this star on a period equal to that of the stellar rotation period.  \\ClearASIL was then applied to these line-profiles and the results are shown in the bottom right panel of Fig.~\ref{fig:1Spot}.   As described in Section 2, ClearASIL fits the signal with a combined planet and spot model, as well as with a spot only model to assess which best fits the data.  For this model star the code found correctly that the best fit for the line-profiles was the spot only fit and was able to effectively clean the true spot features from the line-profiles.   \\The $\chi^2$ levelled off at the 6$^{th}$ iteration and the resulting RVs showed no realistic planet signature.  The resulting RVs had an rms of 7.45$\pm$3~ms$^{-1}$, showing that a 96\% reduction in the stellar noise was achieved.\\

\begin{figure}
\subfloat{\includegraphics[trim = 10mm 10mm 10mm 20mm, clip,width=4cm, height=5.5cm]{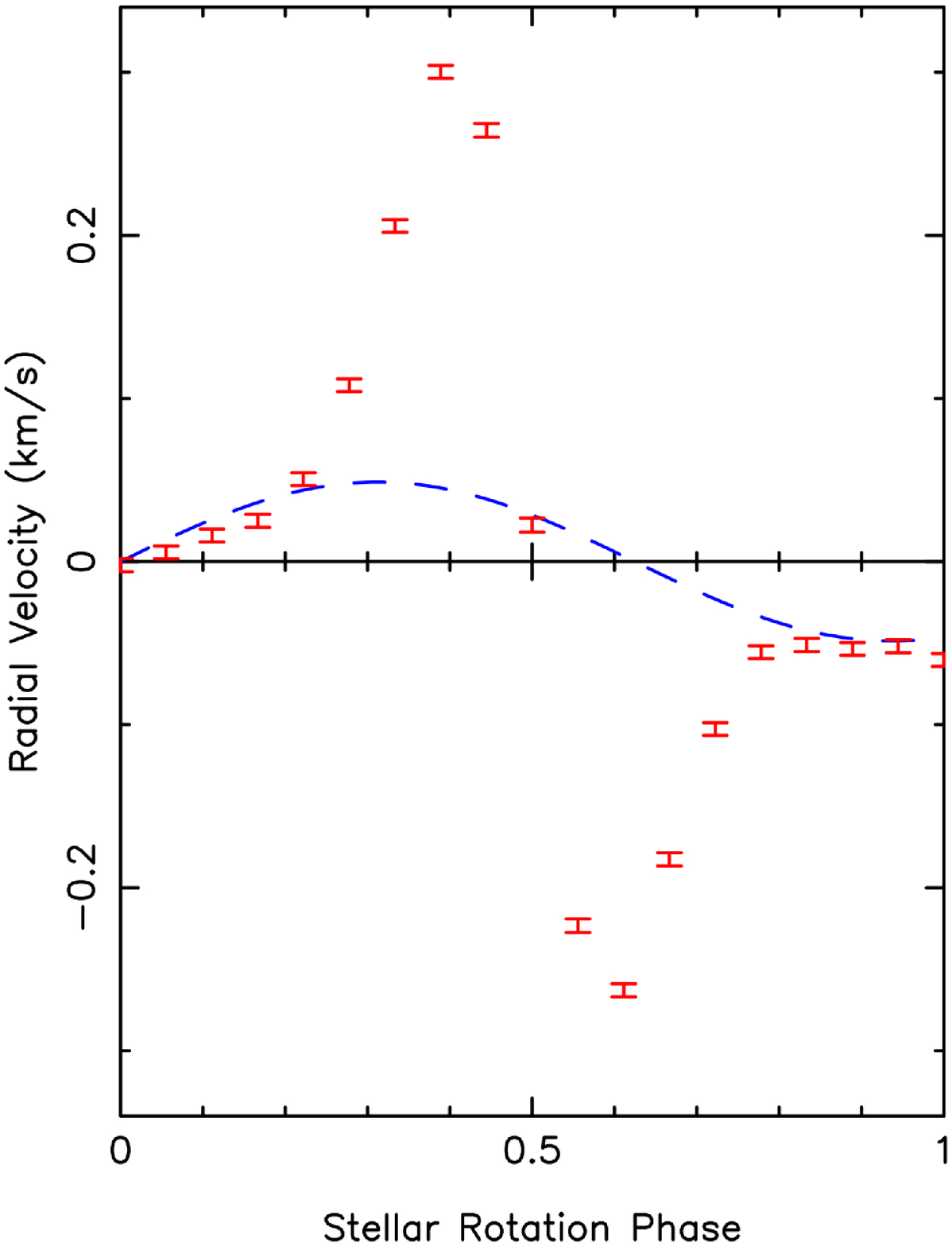}}
\subfloat{\includegraphics[trim = 10mm 10mm 10mm 20mm, clip,width=4cm, height=5.5cm]{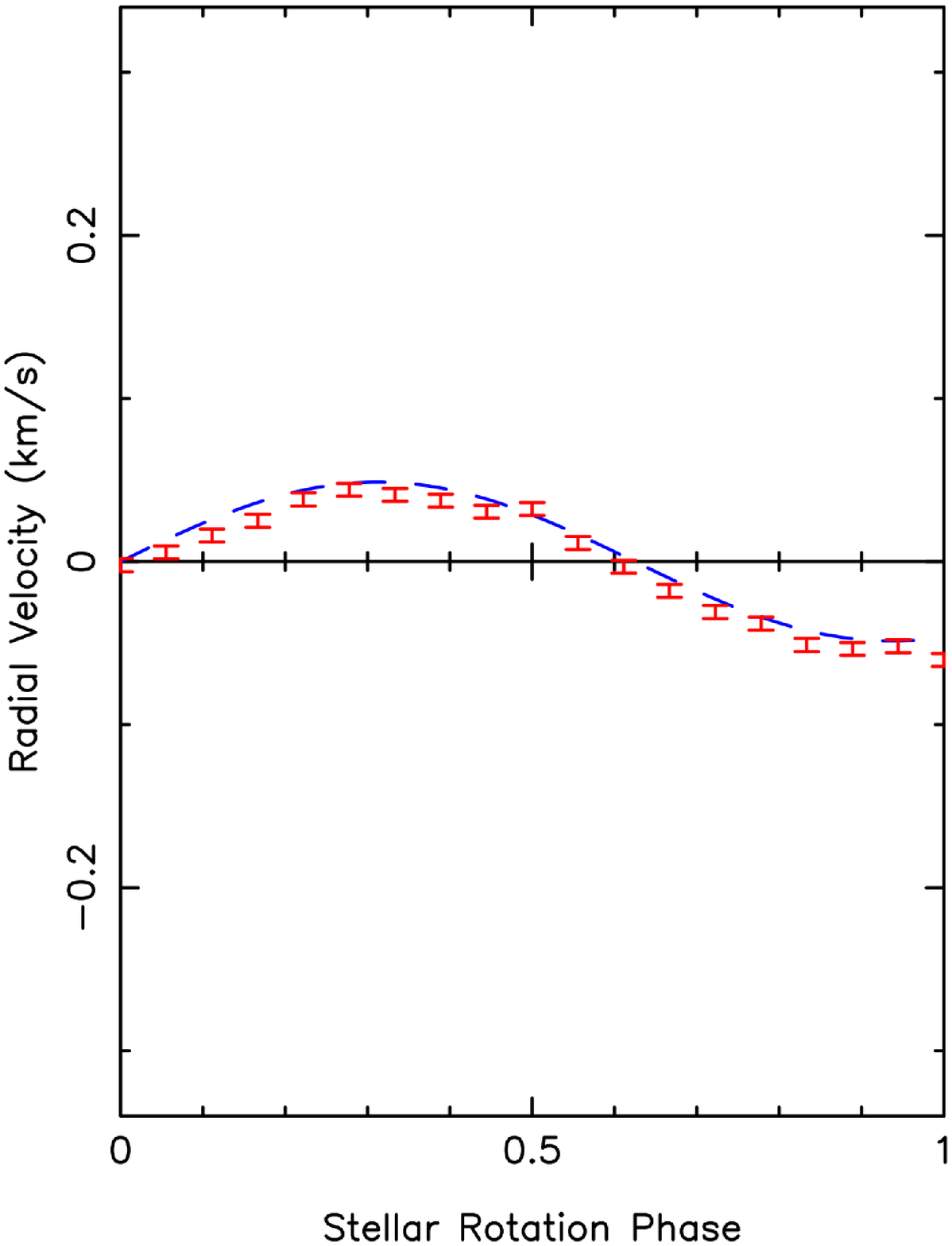}} 
\caption{RV results for the G5V model star which has both a dark spot ($T_{spot}$ = 4600~K) covering 5\% of the visible surface at a latitude of 20$^{\circ}$ and a 1~$M_{J}$ planet on a 4~day orbit.  The crosses represents the RV points of the model data, while the red dashed line corresponds to the RV of the 1~$M_{j}$ planet that has been injected into the model.  The RV of the simulated observational line-profiles is shown in the left panel, while the right panel shows the RVs after using ClearASIL to remove any spot features present.  As can be seen, the injected planet signature is well recovered from the noise. }
\label{fig:1SpotPlanet}
\end{figure}

Next we tested the case when a relatively small planet signature (compared to the stellar jitter) is present.  We placed a 1~$M_{J}$ planet on a 4~day circular orbit around the same model star and again generated 19 line-profiles that evenly covered 1 stellar rotation period.  The RV semi-amplitude caused by this planet is 48.7~ms$^{-1}$, which given the previous test should be discernible if our spot removal algorithm does not overly affect the planet signature.  We note that, in this test, the stellar rotation period and planetary orbital period are similar.  Thus, for a significant fraction of our simulated observations, the RV jitter due to the spot, dominates over the weaker planet signature and is phased similarly (as can be seen in Fig.~\ref{fig:1SpotPlanet}).  \\For the first iteration of ClearASIL the spot only fit was found to best remove any spot features from these line-profiles.  However, in subsequent iterations, the code cleaned the line-profiles using the combined planet and spot fit.  The $\chi^2$ reached a minimum at the 10$^{th}$ iteration and the results showed there to be a periodic RV variation with a semi-amplitude of 49.8$\pm$1.4~ms$^{-1}$ which closely matches the RV signal of the injected planet (see right panel of Fig.~\ref{fig:1SpotPlanet}).    Fitting these RVs with a sine wave reveals a 1.03$\pm$0.05~$M_{J}$ on a 4.08$\pm$0.115~day orbital period, indeed confirming the results are consistent with the same as the planet injected into this model.  This simple test case shows that ClearASIL was not only capable of removing the majority of a large spot from the line-profiles but also has the ability to reveal a planet signal that is partially hidden in the RV data.

\begin{figure}
\subfloat{\includegraphics[width=8cm, height=6cm]{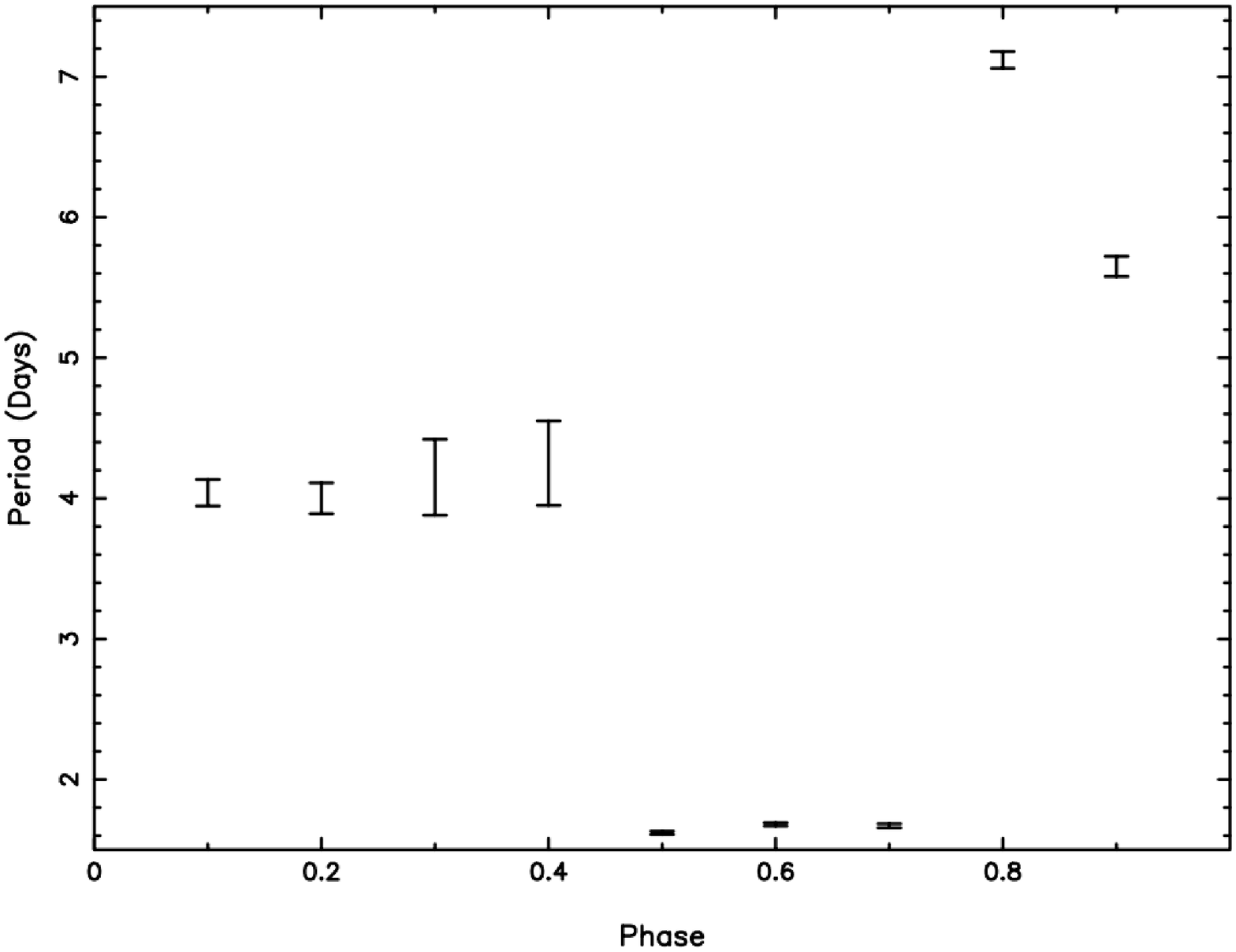}}
\\
\subfloat{\includegraphics[width=8cm, height=6cm]{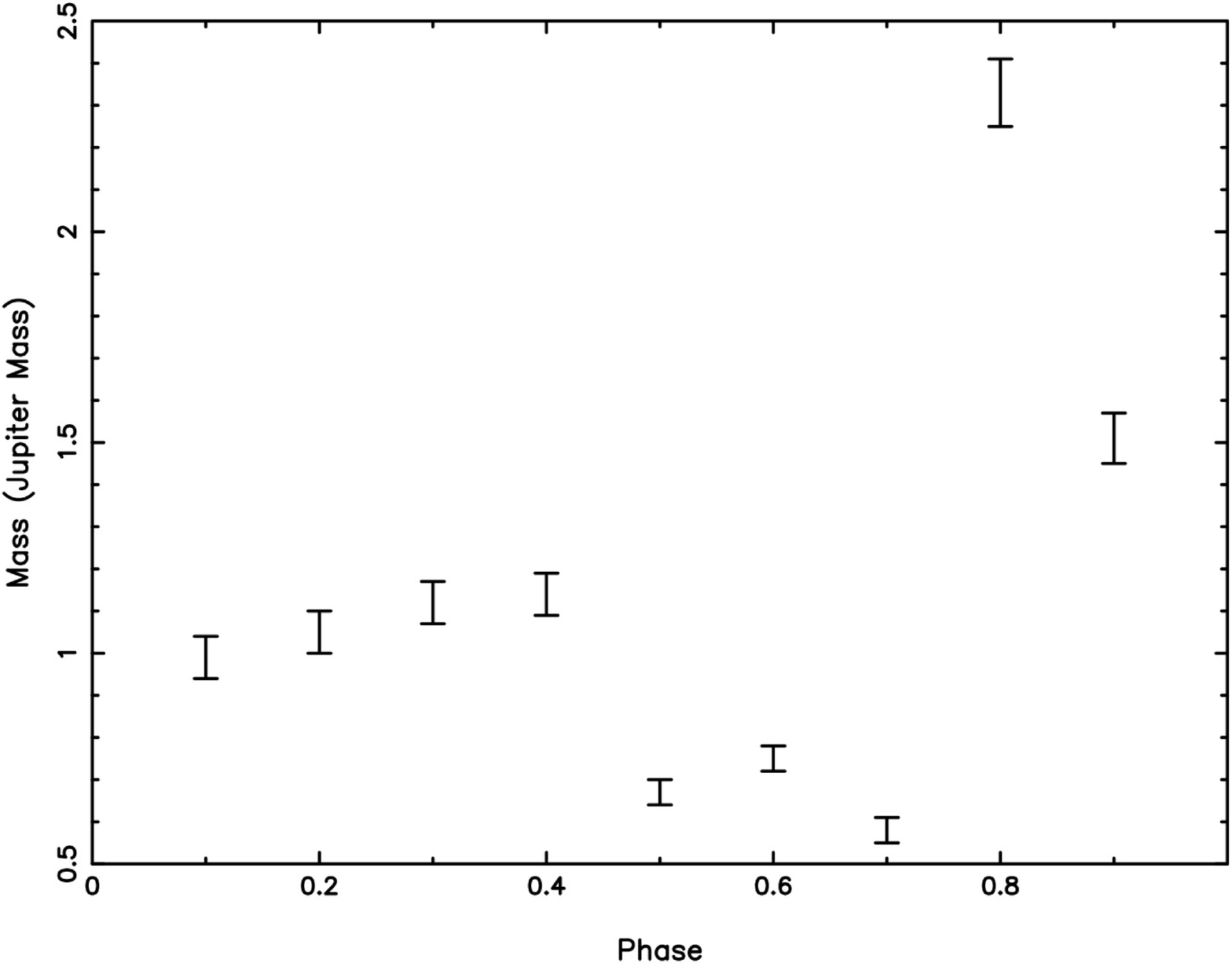}} 
\caption{Results showing the impact of varying the planet signal phase on the ability of ClearASIL to uncover the correct planet signature.  In this case a G5V model star with a dark spot ($T_{spot}$ = 4600~K) covering 5\% of the visible surface at a latitude of 20$^{\circ}$ and a 1~$M_{J}$ planet on a 4~day orbit was used.  The phase of the planet was varied from 0 to 1 in steps of 0.1.  The top panel displays how the planetary period uncovered by ClearASIL varies with the injected planetary phase.  While the bottom panel shows the planetary phase against the mass of the uncovered planet signal.}
\label{fig:1SpotPlanetPhase}
\end{figure}

Having shown that the code was able to impressively uncover the 1~$M_{J}$ mass signal hidden in the stellar noise we decided to further test the ability of ClearASIL.  In the next case we investigated the impact the phase of the injected planet signal had on the planet signature that could be uncovered from the noise.  Again we used the same 1 spot model star with a 1~$M_{J}$ planet on a 4~day circular orbit.  The phase of the planet compared to the stellar rotation phase was varied between 0 and 1 in steps of 0.1.\\  These models were then processed through ClearASIL and the impact the phase had on the mass and period of the planet detected is shown in Fig.~\ref{fig:1SpotPlanetPhase}.  When the phase of the planet is less than 0.5 the code is able to successfully remove the spot signatures to reveal the 1~$M_{J}$ planet with a 4~day orbital period.  However when the planet phase is set to 0.5 or greater the code struggles to uncover the correct planet signal.  This is due to the interplay between the planet signal and the spots causing the code to struggle when fitting the RV data.  Increasing the data size would help the code determing the correct planet signal in cases like this.\\

\begin{figure}
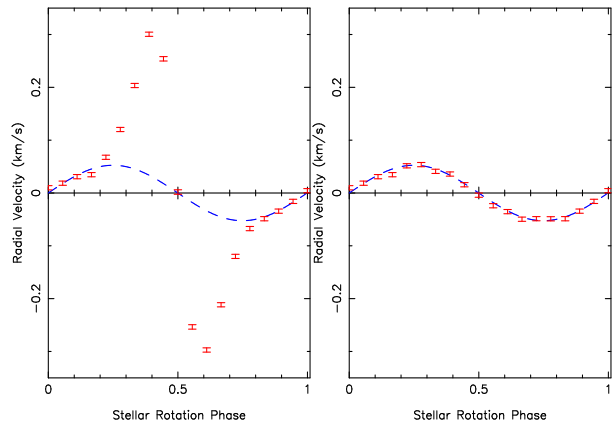

\subfloat{\includegraphics[width=4cm, height=5.5cm]{Moulds_Figure16a.eps}}
\subfloat{\includegraphics[width=4cm, height=5.5cm]{Moulds_Figure16b.eps}} 
\caption{Left panel: RV curve of the G5V model star with a dark spot ($T_{spot}$ = 4600~K) covering 5\% of the visible surface at a latitude of 20$^{\circ}$ and a 1~$M_{J}$ planet on a 3.2~day orbit.  Right panel: RV curve after the spot removal code is applied to the data.  As can be seen ClearASIL is successful at reducing the RV noise to reveal a periodic signal that closely matches the signal of the injected planet.}
\label{fig:1SpotPlanetStellarPeriod}
\end{figure}

This next test case is to show the ability of the code on uncovering a planet signal that is the same as the stellar rotation period.  Again the same model star with 1 spot covering 5\% of the stellar surface was used but this time the injected planet signal had mass of 1~$M_{j}$ and a period of 3.2~days (i.e. the same as the stellar rotation period).  The initial semi-amplitude of the RV signal was 299$\pm$5.1~ms$^{-1}$ which completely masks the RV signal of the planet (see Fig.~\ref{fig:1SpotPlanetStellarPeriod}). \\  The spot removal code was able to successfully reduce the stellar jitter to reveal  a periodic RV variation with a semi-amplitude of 51.5$\pm$1.4~ms$^{-1}$ which closely matches the RV signal of the injected planet (see right panel of Fig.~\ref{fig:1SpotPlanetStellarPeriod}).   Fitting these RVs with a sine wave reveals a 0.98$\pm$0.04~$M_{J}$ on a 3.2$\pm$0.04~day orbital period, indeed confirming the results are consistent with the same as the planet injected into this model.  This highlights the ability of ClearASIL to uncover the correct planetary parameters even if the planetary orbital period is the same as the stellar rotational period.\\    

\subsection[]{4 Spot Model}
\label{sec:4spots}
\begin{figure}
\subfloat{\includegraphics[trim = 10mm 10mm 10mm 20mm, clip,width=4cm, height=5.5cm]{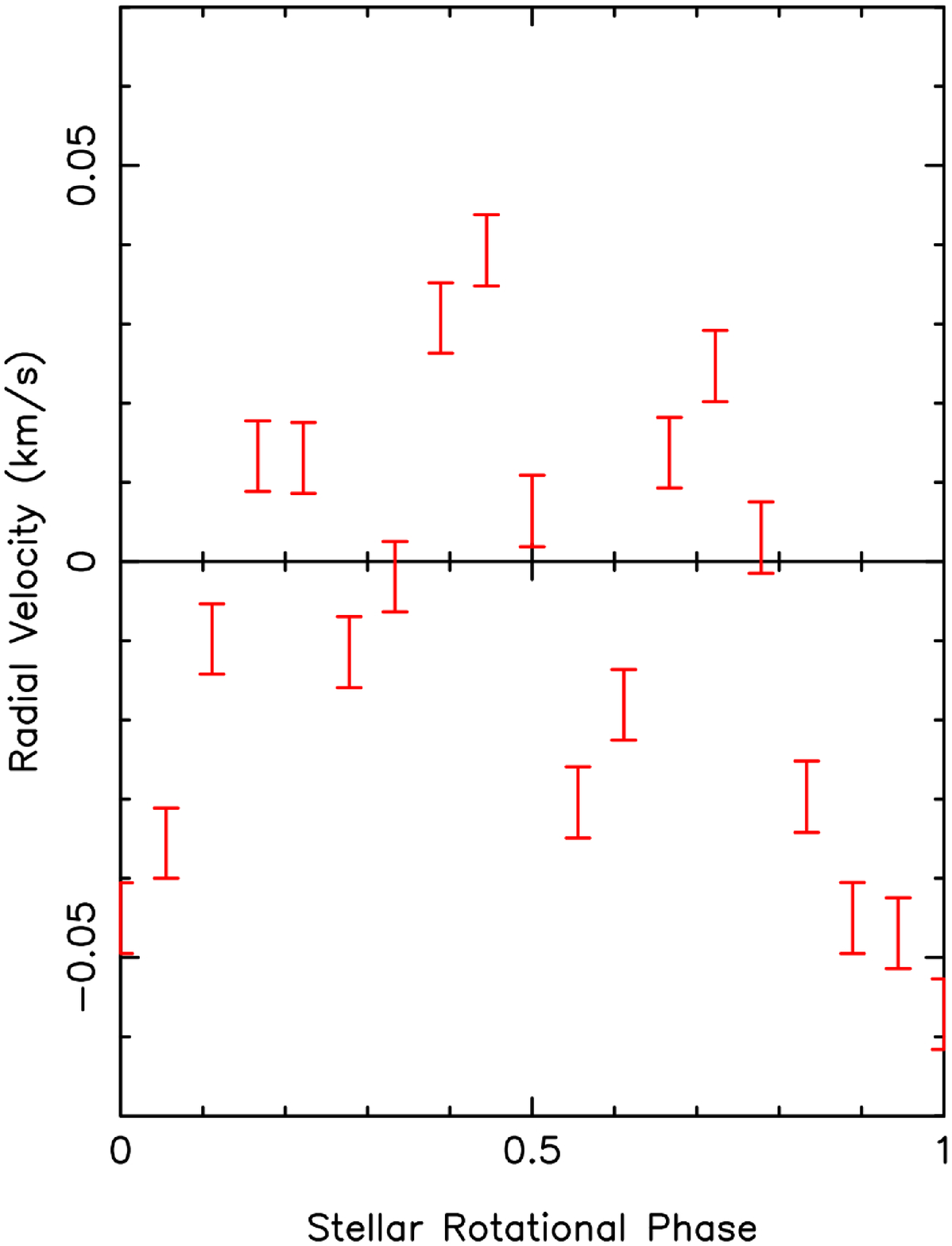}}
\subfloat{\includegraphics[trim = 10mm 10mm 10mm 20mm, clip,width=4cm, height=5.5cm]{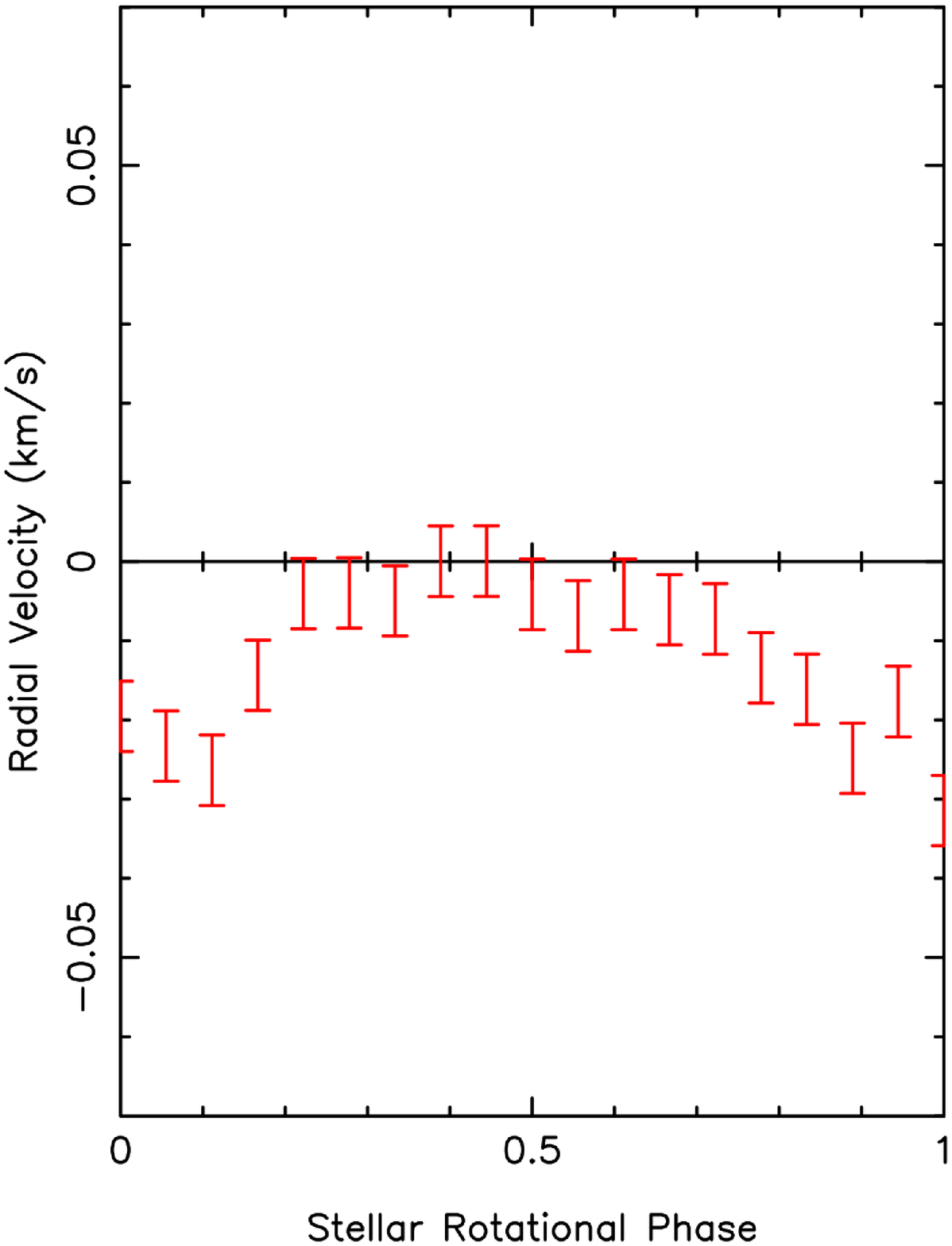}} 
\caption{Left panel: RV curve of the G5V model star with 4 dark spots ($T_{spot}$ = 4600~K) each covering 1\% of the visible surface at positions detailed in Table 1.  Right panel: RV curve after 13 iterations of the spot removal code.  As can be seen ClearASIL is successful at reducing the RV noise by 75\%. }
\label{fig:4Spot}
\end{figure}

The previous 1 spot model represents a simple test case.  In this section we investigate the ability of ClearASIL to cope with a more realistic multiple spot model.   Four spots were placed on the model G5V star (see Section 3) and the details of the latitude, size and temperature of these spots can be found in Table~\ref{tab:4spotmodel}.  As for the previous model, 19 line-profiles were generated covering 1 stellar rotation period.  The RVs resulting from this model are not very periodic due to the interplay of multiple spots crossing the stellar disc and have an rms of 30~ms$^{-1}$ and a peak amplitude of 96~ms$^{-1}$, as shown in Fig.~\ref{fig:4Spot}.  \\The line-profiles were put through ClearASIL, which correctly found the spot only fit to be best for all iterations of the code.  The $\chi^2$ levels out at the 13$^{th}$ iteration and the resulting line-profiles reveal an RV signal with an rms of 13~ms$^{-1}$ (Fig.~\ref{fig:4Spot}).  In this case ClearASIL has effectively reduced the stellar jitter by 57\%.  The peak to peak amplitude is reduced from 96$\pm$6.3~ms$^{-1}$ to 24$\pm$6.2~ms$^{-1}$ which is a 75\% reduction in amplitude.  Although this is a significant result, we do note it is smaller than the RV improvement seen in the 1 spot model case under the same sampling and SNR conditions.  ClearASIL assumes nothing about the number of spots present on the star and fits the line-profiles for any spot-like features according to the fitting technique outlined in Section~\ref{sec:spotremoval}.  There is no limit on the number of spots that ClearASIL will try and fit and remove from the line-profiles.  Therefore if ClearASIL detects 10 spots in the line-profile then it will fit for 10 spots or if it detects 20 spots then it will fit for 20 etc.  Multiple, smaller spots are difficult to fit using this Gaussian based fitting technique and so the line-profiles are not cleaned as effectively in this 4 spot model. \\We also note that in this instance the resultant RVs appear to have a systemic velocity of approximately -13~ms$^{-1}$ which does not match the systemic velocity of 0~ms$^{-1}$ that was given to the model data.  When dealing with multiple spots rotating across the star it becomes more likely that a spot feature will appear in the same place in all the line-profiles.  If this does happen then when the line-profiles are averaged together to create the psuedo-immaculate profile as described in section~\ref{sec:spotremoval} this feature will also appear in the average profile.  Subtracting this psuedo-immaculate profile from each individual profile will have the result of cancelling this spot feature out and so it will never appear in the residuals.  This results in the code being unable to isolate this spot feature and remove it from of the line-profiles.  \\Although this is annoying it does not have any impact on finding planets that could be hidden in the data.  Not removing this spot feature from each of the line-profiles will create the same \textit{apparent} RV shift in all the line-profiles and so will effectively change the systemic velocity.  Any \textit{true} shift in the centre of the line-profiles due to an orbiting planet will still be visible as is shown in the following case when a planet is added to this data.\\    

\begin{center}
\begin{table*}
\caption[]{A description of the size, position and temperatures of the spots placed on the 4 spot model \it{\color{Red}G5V star}.}

\begin{tabular}{ | p{2cm} |  p{2cm} | p{2cm} | p{2cm} | p{2cm} |}
    \hline
    Spot Number & Latitude & Longitude & Size (\% of visible surface) & Temperature in K \\ \hline
    1 & 30 & 0 & 1 & 4600 \\ \hline
    2 & -20 & 90 & 1 & 4600 \\ \hline
    3 & -5 & -90 & 1 & 4600 \\ \hline
    4 & 60 & 120 & 1 & 4600 \\ 
    \hline

\label{tab:4spotmodel}

\end{tabular}
\end{table*}
\end{center}

\begin{figure}
\subfloat{\includegraphics[trim = 10mm 10mm 10mm 20mm, clip,width=4cm, height=5.5cm]{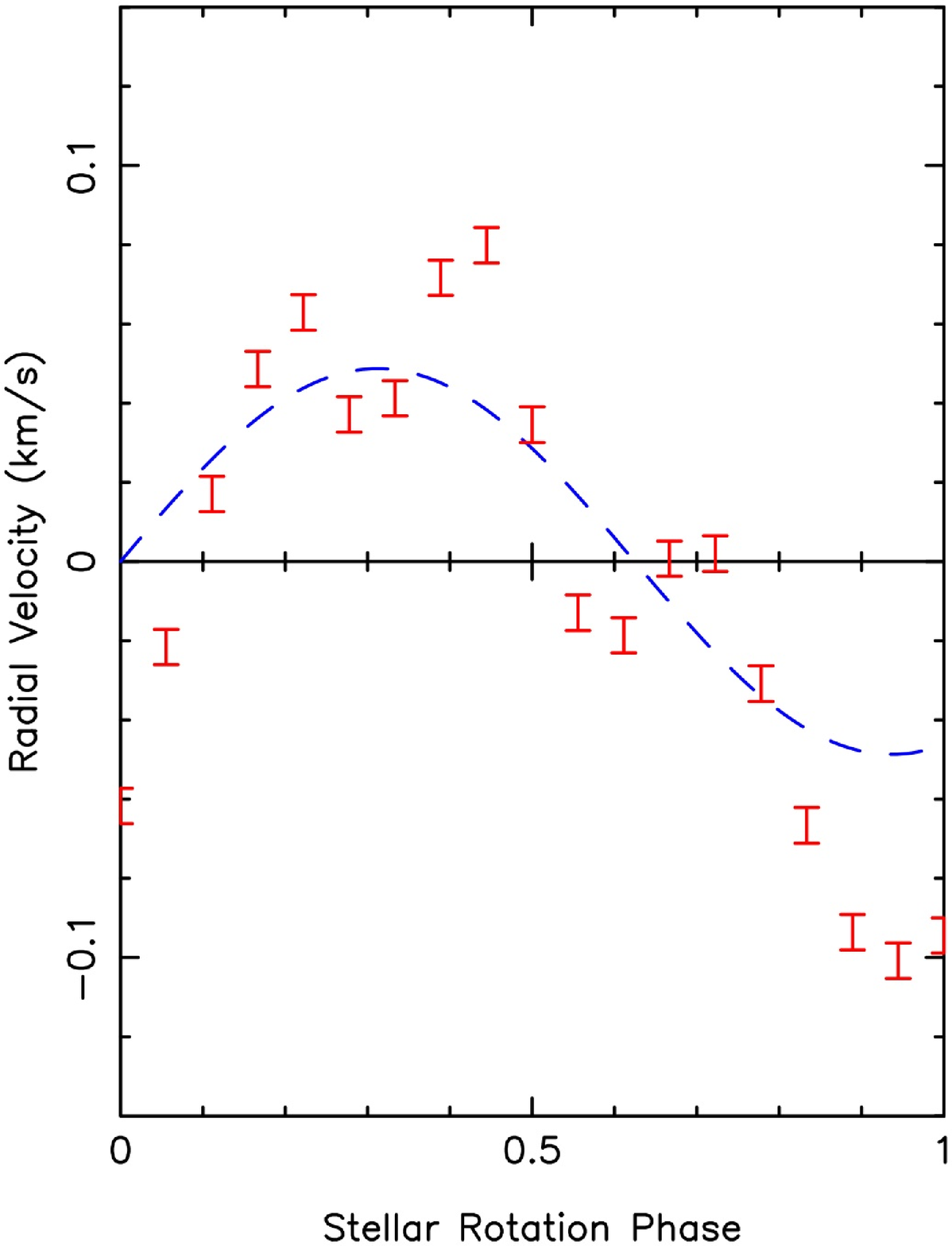}}
\subfloat{\includegraphics[trim = 10mm 10mm 10mm 20mm, clip,width=4cm, height=5.5cm]{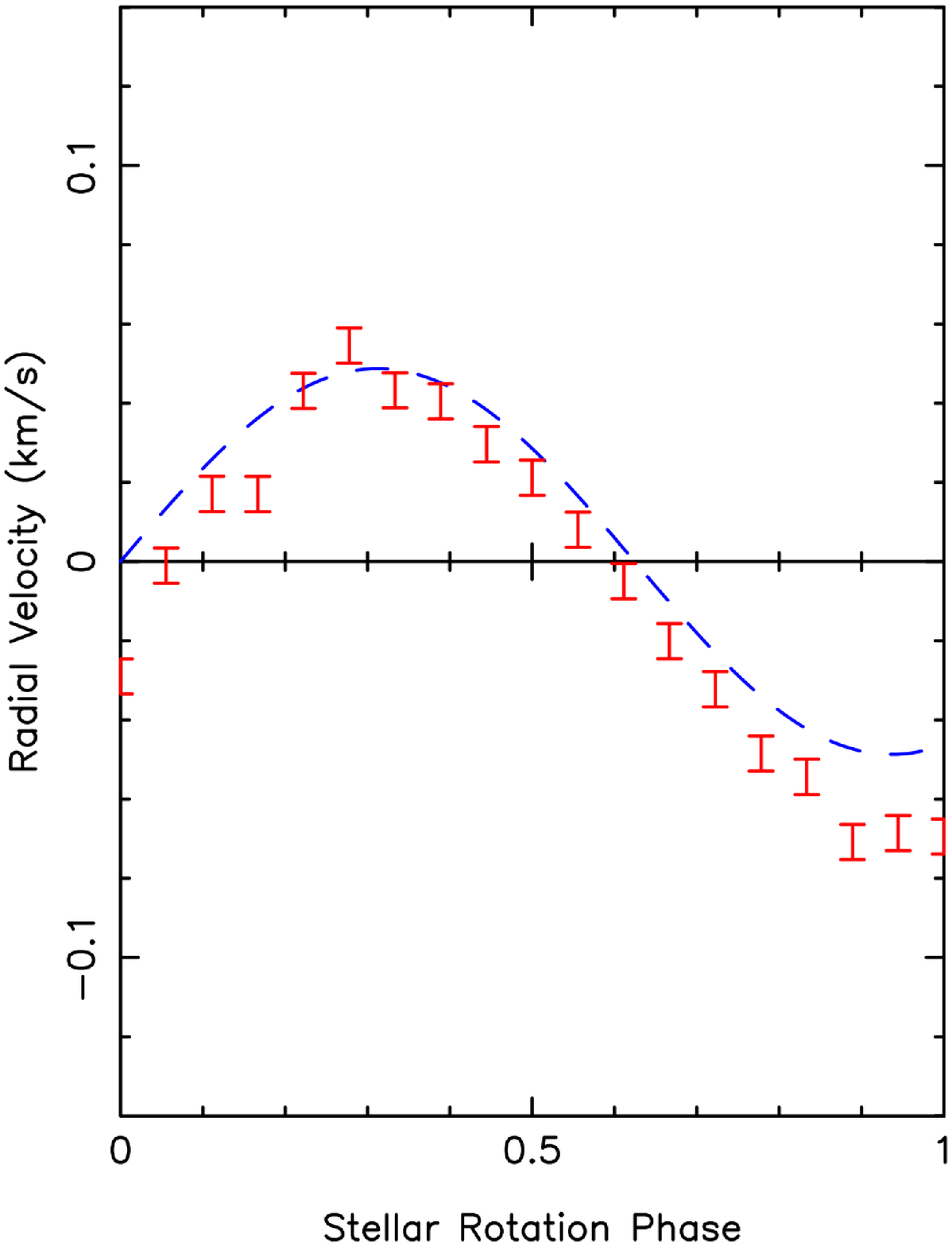}} 
\caption{Left panel: RV curve of the G5V model star with 4 dark spots ($T_{spot}$ = 4600~K) each covering 1\% of the visible surface at positions (detailed in Table 1) and a 1 Jupiter mass planet on a 4~day orbit.  Right panel: RV curve after 21 iterations of the spot removal code.}
\label{fig:4SpotPlanet}
\end{figure}

As with the previous 1 spot case, we decided to test the code further by inserting a 1~$M_{J}$ planet on a 4~day circular orbit into this multiple spot model.  Fig.~\ref{fig:4SpotPlanet} shows the RV curve for this model, which has a semi-amplitude of 87.14$\pm$6.3~ms$^{-1}$.  There is a hint of a planetary signal in the RV jitter due to the fact that both the RV amplitude and orbital period of the planet is similar to that of the spots, and they both have similar phase.  Fitting a sinewave to this original `noisy' data reveals a 1.8 $M_J$ planet on a 4.7 day orbital period.\\  ClearASIL was once again able to successfully reduce the stellar jitter and clean the injected planetary signal.  The lowest $\chi^2$ occurred at the 21$^{st}$ iteration with the resulting RV curve having a periodic shape and semi-amplitude of 63$\pm$6.3~ms$^{-1}$.  Fitting a sine wave to the RVs shows a 1.2$\pm$0.15~$M_J$ planet on a 4.15$\pm$0.083~day orbital period.  This is an improvement in the orbital period by 20\% and the planet mass by 30\%.  Due to the confusion of fitting multiple spots the planet signature is slightly distorted, but again the sine fit reveals a similar mass and orbital period to the fake planet signature injected into the model.

\subsection[]{Variable Spot Model}
\begin{figure}
\subfloat{\includegraphics[trim = 10mm 10mm 10mm 20mm, clip,width=4cm, height=5.5cm]{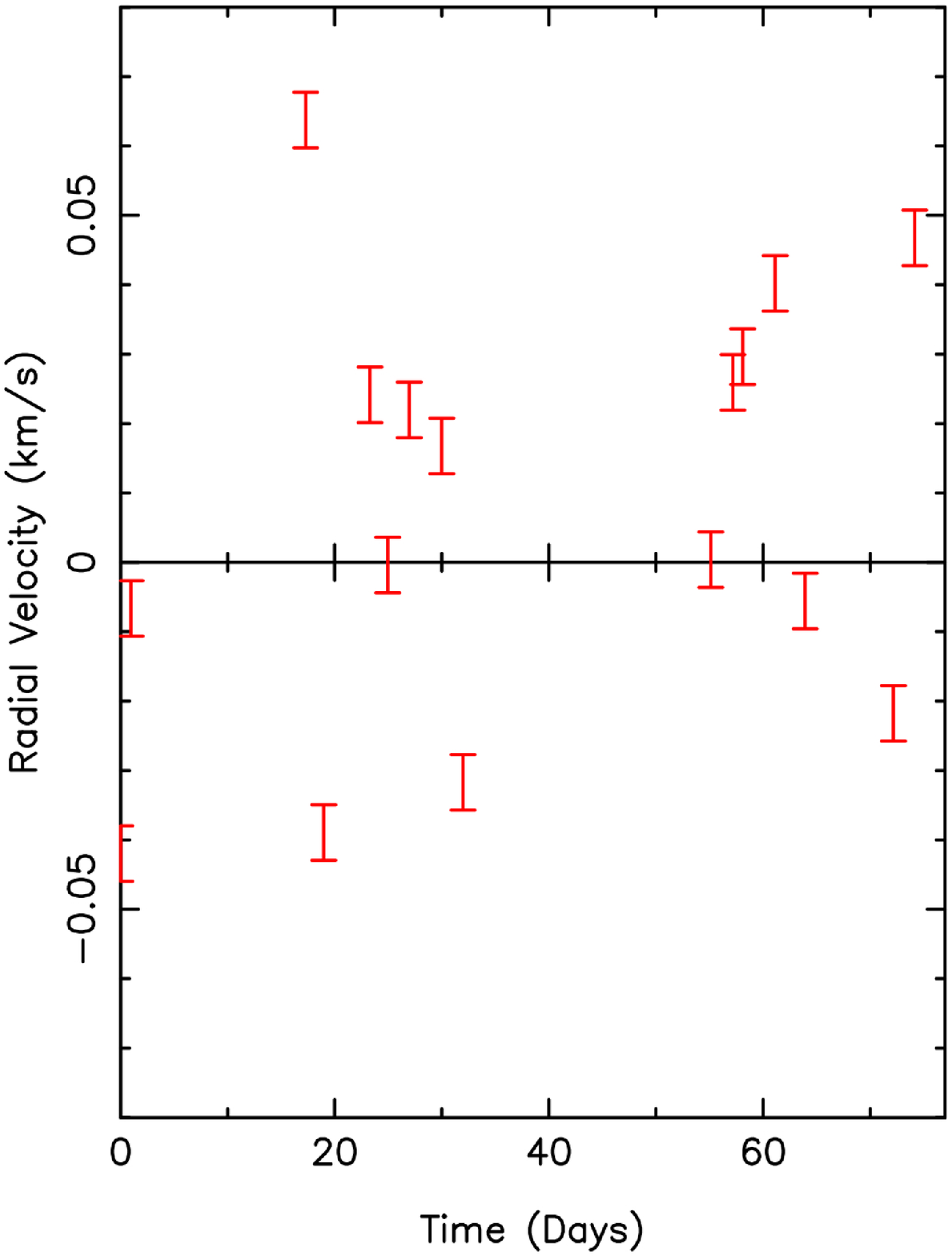}}
\subfloat{\includegraphics[trim = 10mm 10mm 10mm 20mm, clip,width=4cm, height=5.5cm]{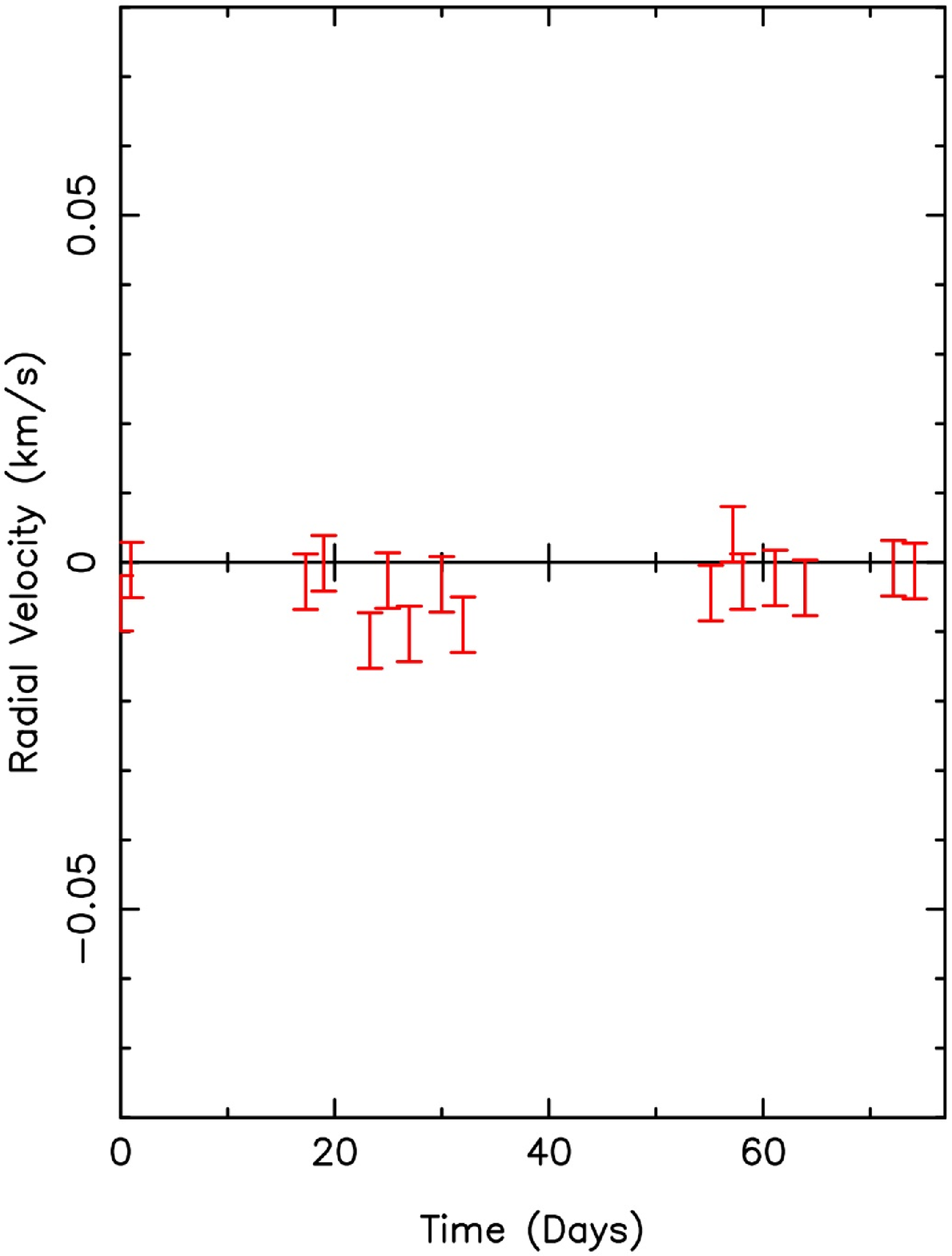}}
\caption{Left panel: RV curve of the G5V model star with random spot coverage ($T_{spot}$ = 4600~K) with positions and sizes detailed in Table 2.  Right panel: RV curve after 18 iterations of the spot removal code.}
\label{fig:MultiSpot}
\end{figure}

The previous models discussed have shown ClearASIL works for single and multiple spot scenarios.  However, in both cases the line-profiles have been evenly spaced over one stellar rotational period.  In reality, observations will be spread out due to limitations caused by telescope availability and the stars' visibility.  So we decided to further test the code on unevenly sampled data.\\  In this case the model star ($M_*$ = 1.05~$M_{sun}$, $R_*$ = 0.95~$R_{sun}$, $T_{eff}$ = 5657~K,  $P_{rot}$ = 3.2~days and $i$ = 51.8$^{\circ}$) was used to generate 16 spectra with varying spot features that covered the time-span of the RV observations taken to confirm the transiting nature of WASP-39b (\citealt{Faedi:2011di}).  Note that we generate one model star for each of the 16 spectra with the spot features set to correspond to that particular observation.  WASP-39b was chosen because it has similar properties to the models so far run in this paper and is therefore a good representation of a typical RV follow-up strategy. The 16 line-profiles span over 76~days with a gap of 20~days in the middle of the observations.\\  We initially placed 3 spots on the star and allowed them to change their size ($\alpha$), latitude ($\phi$) and longitude ($\theta$) but not their temperature (which was kept at a constant 4600~K) for the first 8 spectra.  For the last 8 spectra we changed the spot pattern to 4 spots and again allowed these new spots to change their size, latitude and longitude but kept their temperature at a constant 4600~K.  Table~\ref{tab:multispotmodel} shows the spot properties that were used to construct each model line-profile.  Generally spots have lifetimes that span from several days to several weeks (\citealt{Hall:1994di}) and on stars that are more active than the Sun these features tend to cover a large portion of the star and be found at higher latitudes (\citealt{Donati:1999di}, \citealt{StrassmeierA:1999di}).  I tried to use this information to determine the size and latitudinal position of the spot features on the star as well as for determining how long the spot feature should be present on the surface.  Note the longitudinal position of the spots was choosen so that they would all appear on the surface of the star at the same time.  \\The RVs for this model were found to have an rms of 32.3~ms$^{-1}$, as shown in Fig.~\ref{fig:MultiSpot} and ClearASIL was able to reduce the stellar jitter rms by 65\%.  The peak to peak amplitude was reduced from 105$\pm$6.3~ms$^{-1}$ down to 13$\pm$6.2~ms$^{-1}$ showing an 80\% reduction.  Again we note a systemic velocity present in the resulting data which is once again a result of fitting multiple spots as explained in section~\ref{sec:4spots}.\\

\begin{center}
\begin{table*}
\caption[]{A description of the size and position of the spots placed on the variable spot model G5V star.  The size of the spots is given as the \% of the stellar surface that the spot covers and is denoted by $\alpha$.  The position of the spots are given in terms of their longiture ($\theta$) and latitude ($\phi$) on the stellar surface.  Note the spots have a fixed temperature of 4600~K.}

\begin{tabular}{  | c | c | c | c | c | c |  c | c | c | c | c | c | c |}
\hline
Obs Date (Days) & $\phi$1 & $\phi$2 & $\phi$3 & $\phi$4 & $\theta$1 & $\theta$2 & $\theta$3 & $\theta$4 & $\alpha$1 & $\alpha$2 & $\alpha$3  & $\alpha$4\\ \hline
0 & 30 & -5 & 60 & - & 0 & -90 & 120 & - & 1 & 1 & 1 & -  \\ \hline
0.967 & 30 & -5 & 60 & - & 0 & -90 & 120 & - & 1 & 1 & 1 & - \\  \hline
17.278 & 40 & -5 & 55 & - & 0 & -90 & 120 & - & 1.6 & 1.2 & 1 & - \\ \hline
18.9685 & 40 & -5 & 55 & - & 0 & -90 & 120 & - & 1.6 & 1.2 & 1 & - \\ \hline
23.284 & 60 & -2 & 50 & - & 0 & -90 & 120 & - & 1.2 & 0.9 & 0.9 & - \\ \hline
24.948 & 60 & -2 & 50 & - & 0 & -90 & 120 & - & 1.2 & 0.9 & 0.9 & - \\  \hline
26.9597 & 60 & -2 & 50 & - & 0 & -90 & 120 & - & 1 & 0.7 & 0.85 & - \\  \hline
29.992 & 60 & -4 & 45 & - & 0 & -90 & 120 & - & 0.75 & 0.5 & 0.6 & - \\ \hline
31.9773 & 60 & -4 & 45 & - & 0 & -90 & 120 & - & 0.75 & 0.5 & 0.6 & - \\  \hline
55.1228 & 20 & -5 & 40 & 10 & -110 & 30 & 150 & -25 & 1 & 1.2 & 0.5 & 2 \\ \hline
57.1848 & 20 & -5 & 40 & 10 & -110 & 30 & 150 & -25 & 1 & 1.2 & 0.5 & 2 \\ \hline
58.0972 & 25 & 10 & 40 & 10 & -110 & 30 & 150 & -25 & 1.1 & 1.2 & 0.8 & 1.8 \\ \hline
61.1231 & 25 & -10 & 45 & 20 & -110 & 30 & 150 & -25 & 1.5 & 0.9 & 0.9 & 1.4 \\ \hline
63.9251 & 25 & -10 & 45 & 20 & -110 & 30 & 150 & -25 & 1.5 & 0.9 & 0.9 & 1.4 \\\hline
72.169 & 30 & -5 & 50 & 15 & -110 & 30 & 150 & -25 & 2 & 0.4 & 1.3 & 0.8 \\ \hline
74.1561 & 30 & -5 & 50 & 15 & -100 & 30 & 150 & -25 & 2 & 0.2 & 1.2 & 0.5\\  \hline
76.167 & 30 & -5 & 50 & 15 & -110 & 30 & 150 & -25 & 2 & 0.2 & 1.2 & 0.5\\ \hline

\label{tab:multispotmodel}

\end{tabular}
\end{table*}
\end{center}

\begin{figure}
\centering
\includegraphics[width=6cm, height=6.5cm]{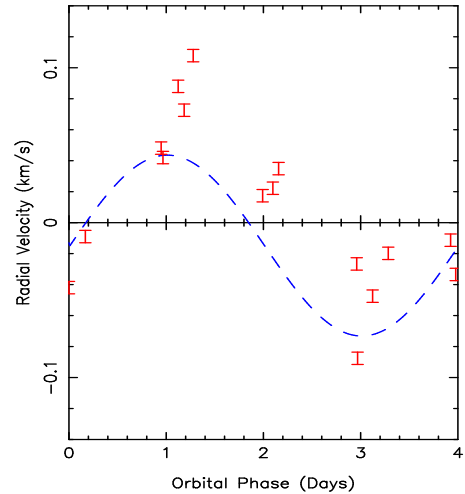}
\caption{RV curve of the G5V model star with random spot coverage (spot parameters are detailed in Table 2) and a 1 Jupiter mass planet on a 4~day orbit.  Overplotted on this curve is a blue dashed line showing the best sine fit to this `noisy' data.}
\label{fig:PlanetFitOrigData}
\end{figure}

\begin{figure}
\subfloat{\includegraphics[trim = 10mm 10mm 10mm 20mm, clip,width=4cm, height=5.5cm]{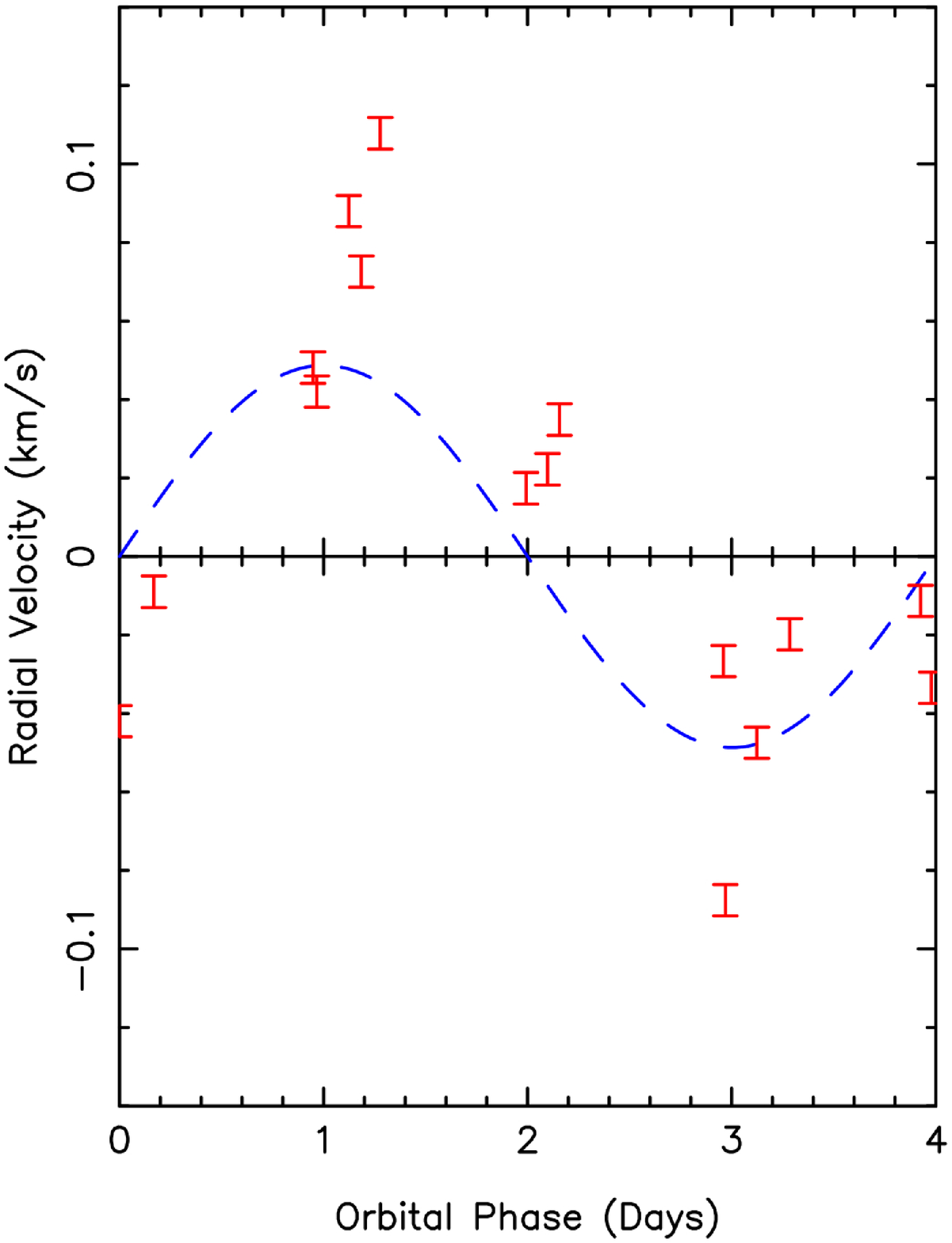}}
\subfloat{\includegraphics[trim = 10mm 10mm 10mm 20mm, clip,width=4cm, height=5.5cm]{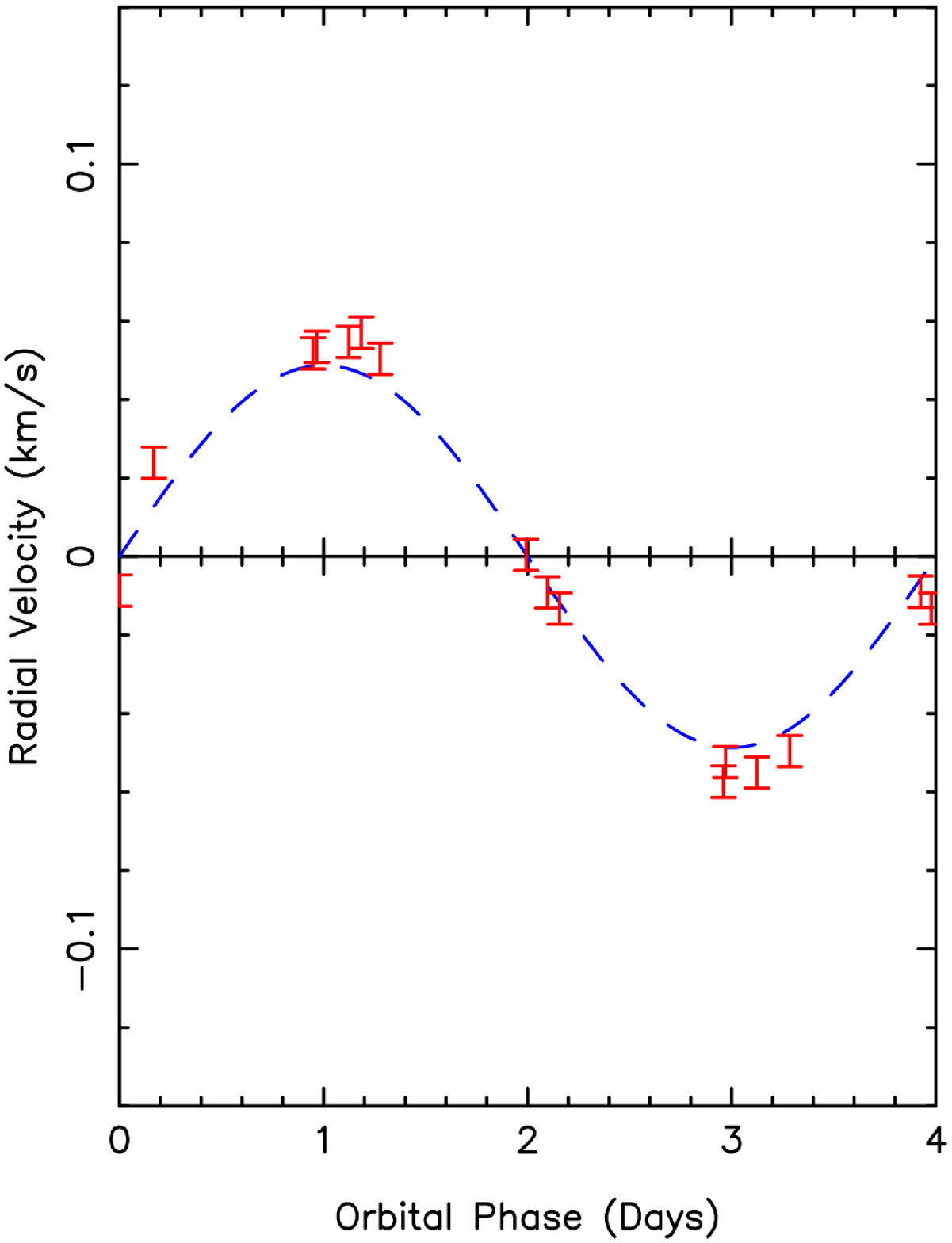}}
\caption{Top: RV curve of the G5V model star with random spot coverage (spot parameters are detailed in Table 2) and a 1 Jupiter mass planet on a 4~day orbit.  Bottom: RV curve after 9 iterations of the spot removal code.  In both plots the data has been phase folded onto the orbital period of the fake planet injected into the data. }
\label{fig:MultiSpotPlanetResult}
\end{figure}

Again we carried out the test when a 1 Jupiter mass planet on a 4~day circular orbit was inserted into this model.  Fig.~\ref{fig:MultiSpotPlanetResult} shows the RVs for this model with the input planet signal over-plotted as the dashed red line.  The RV semi-amplitude due to the planet and evolving spots is 97.5$\pm$6.3~ms$^{-1}$.  Fitting this original `noisy' data with a sine wave reveals a planet with a mass of 1.2~$M_J$ on an orbital period of 4.01~days.  Although this sine wave is close to the true signal it does not match well with the datapoints as shown in Fig.~\ref{fig:PlanetFitOrigData}. \\Even though the data size was small and covered a large time-span, ClearASIL was still able to detect that the planet and spot fit was better at removing the spot from the line-profiles.  The lowest $\chi^2$ was reached at the 9$^{th}$ iteration and the resulting RVs had a semi-amplitude of 45$\pm$6.3~ms$^{-1}$.  Fitting these RVs with a sine wave revealed a 1.14$\pm$0.18~$M_j$ on a 4$\pm$0.08~day orbital period, which is consistent with the planet signature initially placed into the model. The code can still successfully detect the presence of a planet with sparse data covering a large time-span.\\

\section[]{HD49933}
\label{realResults}
\begin{figure}
\subfloat{\includegraphics[trim = 10mm 10mm 10mm 20mm, clip,width=4cm, height=5.5cm]{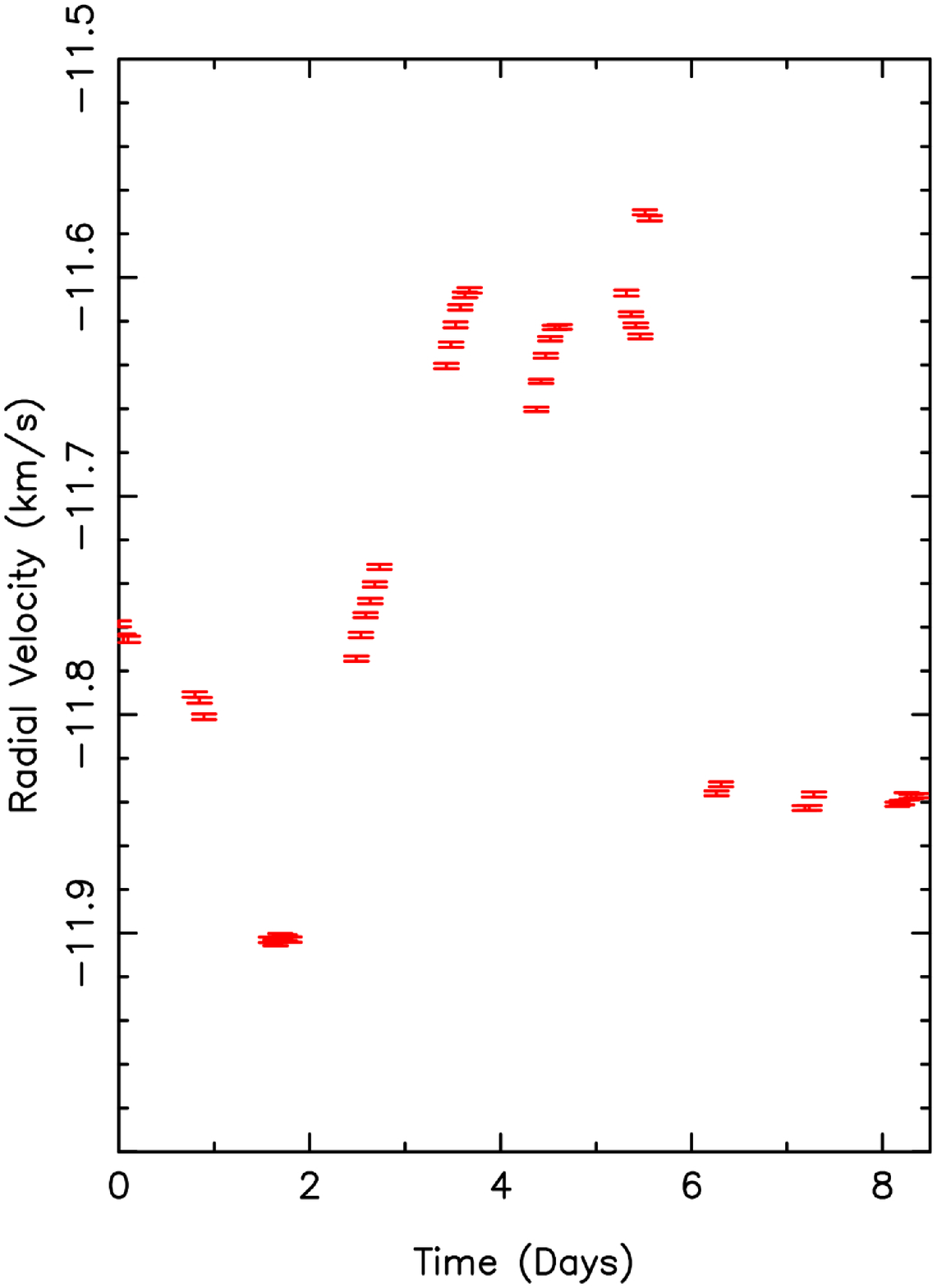}}
\subfloat{\includegraphics[trim = 10mm 10mm 10mm 20mm, clip,width=4cm, height=5.5cm]{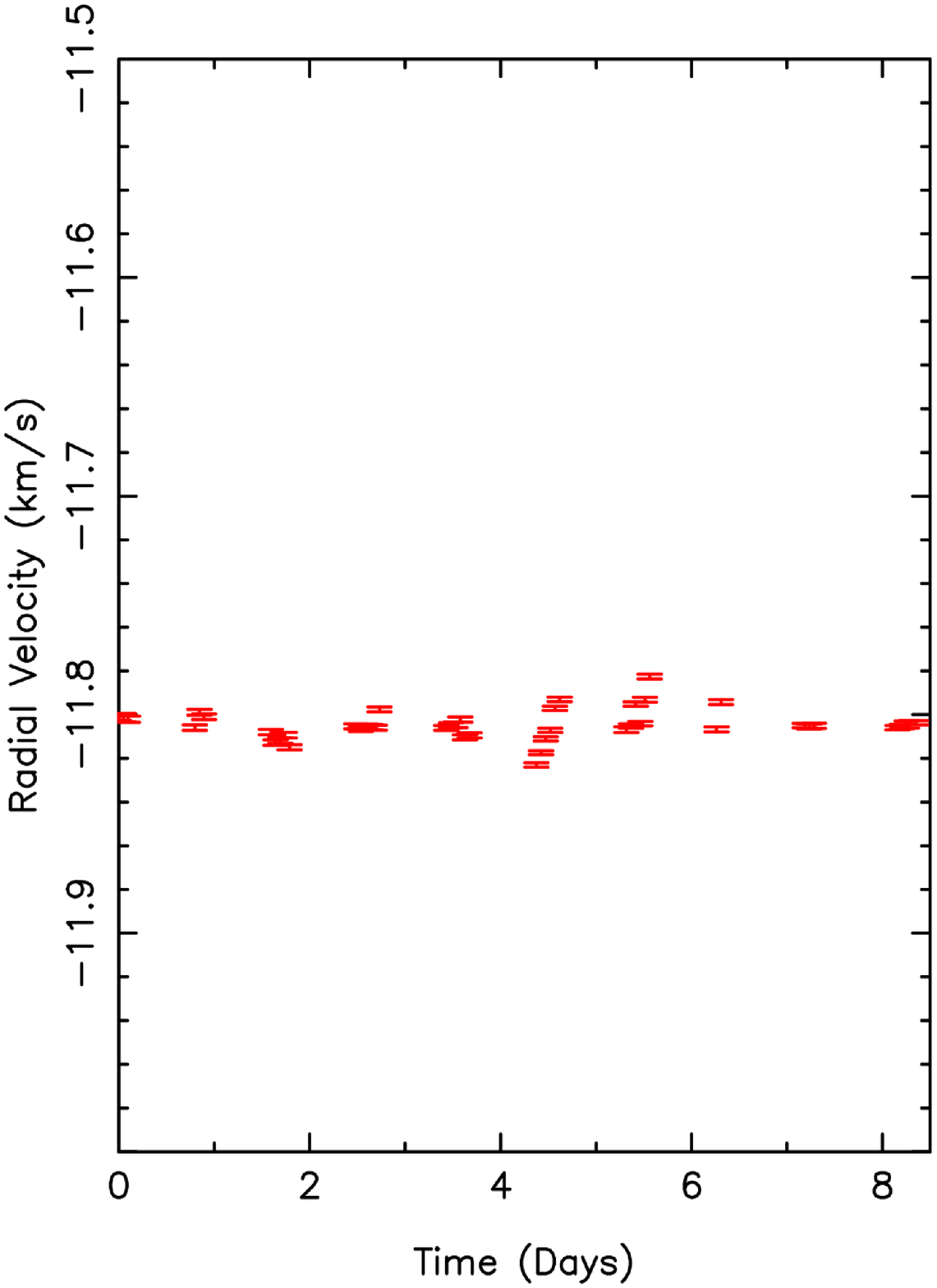}}
\caption{Left panel: RV curve of HD49933.  Right panel: RV curve after the spot removal code.}
\label{fig:RealStar}
\end{figure}

Having tested ClearASIL on model data we now turn to observational data of the F5V main sequence star, HD49933.  HD49933 is known to be magnetically active and has a $v\sin i$ of approximately 10~kms$^{-1}$ making it an ideal candidate for testing the abilities of ClearASIL.  It was observed spectroscopically in 2004 for 10 nights by \cite{Mosser:2005di} who found the star to have large amplitude RV variations as a result of stellar activity.  Later photometry taken by the CoRoT team in 2007 and 2008 (\citealt{Benomar:2009di}; \citealt{Appourchaux:2008di}) showed active regions crossing the stellar disk with a rotation period of 3.5~days.\\We analysed 10 days of archival HARPS data from the observing run of \cite{Mosser:2005di}.  This data set consisted of a total of 1304 spectra which were then processed using the LSD method to produce line-profiles with a significantly higher SNR.  The resulting LSD line-profiles were further binned into batches of 1hr 10mins over each night.  This time was chosen because it corresponded to the star rotating by approximately 5 degrees preventing any spot features from being smeared.  This left us with a total of 45 line-profiles covering 8.5~days.  \\The RV results for this data are shown in Fig.~\ref{fig:RealStar}.  The left plot shows that the data had a peak to peak amplitude of 334.2$\pm$2.1~ms$^{-1}$ initially.  After these line-profiles were processed through ClearASIL the peak to peak amplitude was reduced to 61$\pm$2.2~ms$^{-1}$.  This is a reduction in the stellar jitter of approximately 80\%, comparable to the results from our simulations.  

\subsection[]{HD49933 With Fake Planets}
\begin{figure}
\includegraphics[trim = 10mm 0mm 0mm 40mm, clip,width=9cm, height=8cm]{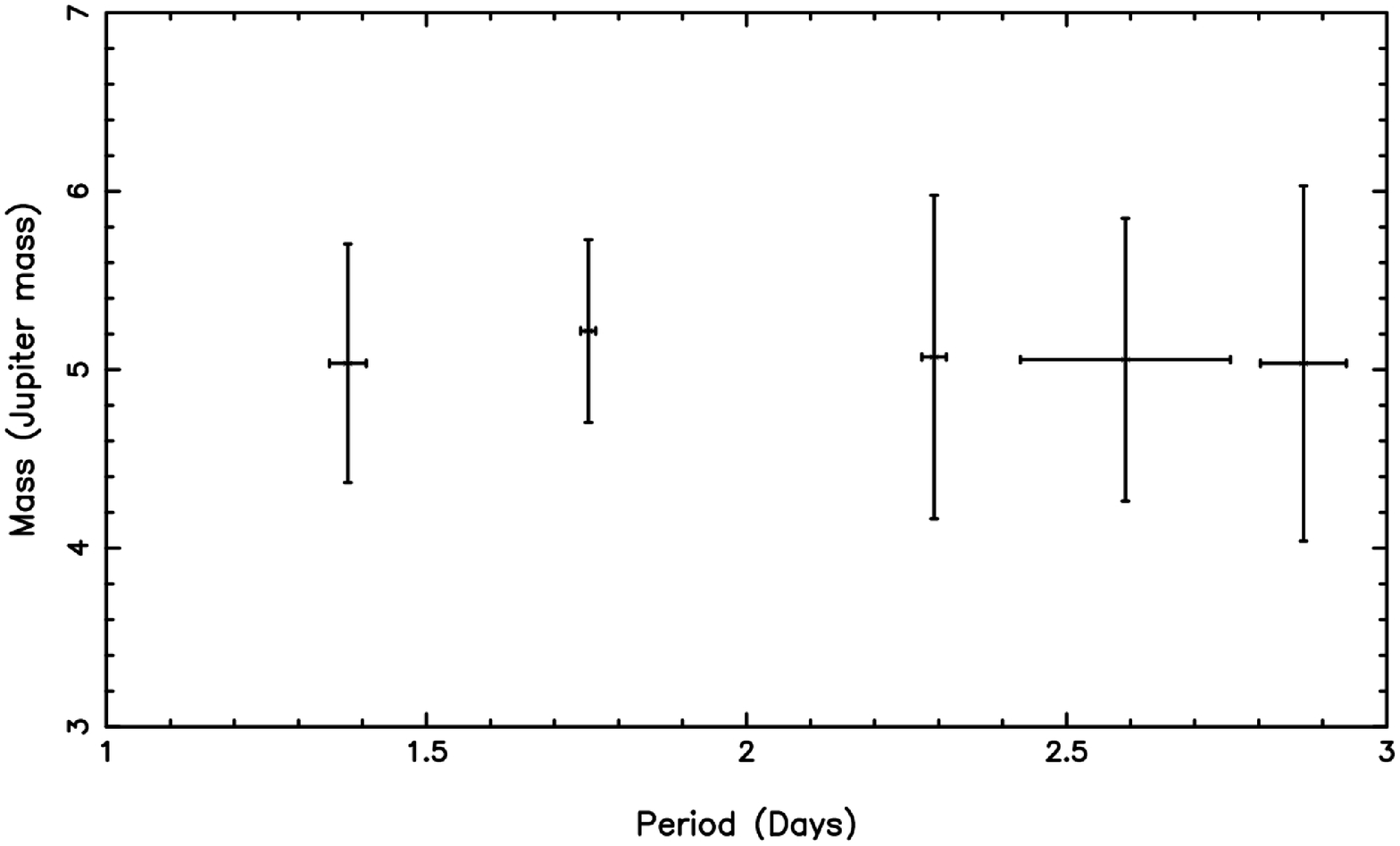}
\caption{The mass of a planet as a function of the planetary period when using ClearASIL to uncover approximately 5 Jupiter mass planets injected into the HD49933 data. The error bars represent the difference between the fitted and the simulated parameters.}
\label{fig:RealStarPlanets}
\end{figure}

Currently HD49933 has not been found to host any planets as has been confirmed by our results in the previous section.  In order to test how a planet would affect the removal of spots with ClearASIL we again injected fake planets with a range of orbital periods into the HD49933 data.  The size of the planet was set to $\sim$5 Jupiter mass. This should be easily detectable by our code which was able to reduce the stellar jitter by 80\%.  The orbital period of the planets were limited to less than 3~days due to the data only spanning 8.5 nights. \\Fig.~\ref{fig:RealStarPlanets} shows the period and mass of planets found by the code in comparison to the model planets injected into the data.  The error bars in this plot are the difference between the injected planet signals and the planets found when using ClearASIL on the data.  They reveal the code to be more accurate for short period planets than for longer period planets which is to be expected due to the short time span of the data.  The plot also displays some scatter for the planet mass showing that the code struggles to get the right planet size.  This is simply due to the star being very active with multiple spots at any one particular time, making it difficult for ClearASIL to accurately fit out all the spot features.  These problems could be overcome with more data points.  Increasing the number of data points improves the average psuedo-immaculate profile enabling any spot features present to be more accurately isolated and removed.  \\To show that this is the case we tested the code when a 1 Jupiter mass planet on 2 day orbital period was injected into the original HD49933 data.  The results from ClearASIL revealed a 2$\pm$0.02 Jupiter mass planet on a 2.64$\pm$0.132 orbital period which differed from the model planet.  To test the effect of increasing the data over a longer time span we doubled the original data points to provide 90 line-profiles and randomly sampled these profiles over 20 days.  Again we injected a 1 Jupiter mass planet on a 2 day orbital period into this new data sample.  ClearASIL was now able to improve on the spot fitting process and the results revealed a 1$\pm$0.05 Jupiter mass planet on a 2.04$\pm$0.01 day orbit.  \\These findings are similar to the model results indicating that ClearASIL could be a powerful tool for uncovering planets distorted or hidden by stellar jitter.           

\section{Comparison with other Techniques}
ClearASIL is not the first technique to remove stellar jitter from planet signals, there are a variety of methods currently in use such as pre-whitening techniques (\citealt{Queloz:2009di}), Fourier Analysis (\citealt{Hatzes:2010di}) and Harmonic decomposition (\citealt{Boisse:2011di}).  However, ClearASIL differs greatly from these methods in the approach it takes at removing stellar jitter.  Instead of analysing the RV data to identify and remove spurious RV signals due to activity, ClearASIL analyses and removes spot features from the actual line-profiles themselves.  This makes it a unique technique that compliments these other methods.  \\Although ClearASIL differs in the method it uses to remove spot features, it is still able to produce results comparable to these other techniques.  \cite{Boisse:2011di} showed that by using harmonic decomposition to fit 3 sine waves at the rotational period of the star and its first two harmonics $P_{rot}$/2 and $P_{rot}$/3, stellar jitter could be reduced by up to 90\%.  This is similar to the results of our work where removal of spot features from line-profiles was also able to reduce the stellar jitter by 90\% in a number of cases.  \\ClearASIL is also not hampered by the same limitations imposed when analysing RV points alone.  One constraint required for successful removal of stellar jitter from RV data is the accurate measurement of the rotational period of the star.  This results in these methods requiring numerous data points covering more than one rotational period of the star.  Unlike methods which analyse the RV data, ClearASIL does not need to know the stellar rotational period and therefore does not require data covering a long time span.  This was shown to be the case in the variable spot model where data is unevenly sampled over 76 days and spot evolution is allowed with ClearASIL still being able to reduce the stellar jitter by 80\%.  \\RV fitting techniques also struggle to reveal planets with orbital periods that are close to the stellar rotational period.  However, as described in Section~\ref{modelResults} ClearASIL recovered planets on a 3.2 day period which is the same as the 3.2 day stellar rotation period.  \\There is a further limitation of the harmonic decomposition technique in that it only works for planets with an RV semi-amplitude that exceeds $\frac{1}{3}$ of the semi-amplitude of the activity signal (\citealt{Boisse:2011di}). In our 1-spot model (Section~\ref{1SpotModel}) ClearASIL was actually able to recover a planetary signal that was $\frac{1}{5}$ of the size of the stellar jitter.  \\Although our technique is not hindered by the same requirements as the aforementioned stellar jitter removal methods, it does have some limitations.  In order for the spot to be resolved in the line-profiles we need the star to have a $v$sin$i$ higher than 10 kms$^{-1}$.  Below this level the $v$sin$i$ becomes lower than the spectrograph resolution and so any change in shape in the stellar lines by spots only results in them being shifted after convolution with the instrumental PSF (as shown by \citealt{Desort:2007di}).  Fitting sine waves to RV data does not depend on the $v$sin$i$ of the star and so is able to work on stars with low $v$sin$i$'s e.g. \cite{Boisse:2011di} who used harmonic decomposition to reduce the stellar jitter in HD189733 which has a $v$sin$i$ of 1 kms$^{-1}$.  Our technique also requires the $v$sin$i$ of the star to be lower than 50 kms$^{-1}$.  Above this level it becomes more difficult to accurately measure the RV due to broadening of the stellar lines.  In order for the technique to be successful, high SNR spectra are needed and thus bright targets are required.\\ 

\section{Future Work On ClearASIL}
ClearASIL is an effective method for distinguishing between stellar activity and planets as shown from the results in Sections~\ref{modelResults} ~\ref{realResults}.  However, there are further improvements that can be made to this technique.  \\Currently, the code uses a model line list for the LSD procedure.  The accuracy of line lists to model real spectra depends upon the precision of the model atmospheres to obtain the wavelengths and depths of intermediate-strength lines.  However, as \cite{Barnes:2004di} pointed out, these model atmospheres become less accurate for cooler stars and so the line list differs more from the true spectrum which can cause problems such as continua sloping in the deconvolved profile.  A way to overcome this issue would be to use a template spectrum of a slowly rotating star with the same spectral type as the target for the deconvolution.  \\These high SNR line-profiles are then used to isolate spot features by subtracting the average line-profile from each individual profile.  By averaging the line-profiles any features in each individual line-profiles are smeared out to generate a `pseudo-immaculate' line-profile.  A better method to generate an immaculate line-profile would be to produce a model rotationally broadened line-profile for the star using its $v$sin$i$.  No spots would be present in this model line-profile and so subtracting it from the individual line-profiles would enhance the isolated spot features.  By improving the accuracy of the modeling and removal of any spots present the code would be able to solve the systemic velocity shift problem mentioned in Section~\ref{sec:4spots}.  \\Most importantly for the application of our technique to real world data, the model planet fit assumes that only one planet is present in the data and that this planet has zero orbital eccentricity.  As more and more planets are being discovered it is becoming apparent that stars are host to multiple planets as well as planets on eccentric orbits.  To take this into account the planet fitting method could also model eccentric orbits as well as multiple planets.

\section{Conclusions}
We have described a new technique for removing stellar jitter in order to aid the discovery of planets around active stars.  Clearing Activity Signals In Line-profiles (ClearASIL) uses high SNR line-profiles to isolate and remove spot features directly from the line-profiles themselves, enabling any planet signatures that are present in the data to be uncovered.  The ability of this method to remove stellar jitter was tested on model data with varying spot features.  The results showed that ClearASIL was able to remove over 80\% of the stellar jitter, similar to techniques such as harmonic decomposition \cite{Boisse:2011di}.  1 Jupiter mass planets were then injected into these models to test whether the code could accurately remove the stellar jitter while revealing the injected planet signal.  The injected planets that were completely hidden by stellar noise and on a similar rotational period to the star were successfully uncovered.  Even in the case when the planet RV signal closely matched that of the spots, ClearASIL was still able to remove the stellar jitter and reveal the planet signal.  \\ClearASIL was also able to cope with spot evolution and unevenly sampled data, reducing stellar jitter by 80\% in the case of no planet.  It was also able to recover the 1 Jupiter mass planet injected into this model where data was sparse and the time span of the observations was large.\\The known active F5V star, HD49933, which had a $v$sin$i$ of 10 kms$^{-1}$ was a perfect test case for our code.  Using the archive HARPS data taken in 2004 by \cite{Mosser:2005di} we were able to reduce the stellar jitter by 80\%.  As this object has no known planets in orbit we injected a 5 Jupiter mass planet into the data to show the code's ability to uncover planet signals at various orbital periods.  The code was able to uncover the planet signal however the accuracy of the planet parameters varied in relation to the orbital period.  As the orbital period of the planet was reduced, the accuracy of the planet parameters established by ClearASIL improved.  This is expected due to the size and sampling of the data set.  Further work is being undertaken to fully explore the parameter space this technique can cover e.g. what $v$sin$i$ works best, the effect of planet mass and different spot models on the accuracy of the results etc.  \\However this current paper shows that this technique does work and can obtain results with only one major requirement and that is the $v$sin$i$ of the star needs to be within 10 to 50~kms$^{-1}$ in order to be able to resolve the spot in the stellar line-profiles and accurately measure the RVs.  ClearASIL does not require any prior knowledge about the star or a high time series of evenly sampled data, thus making it complimentary to other stellar jitter removal techniques such as harmonic decomposition which do have these limitations but can operate for stars with low $v$sin$i$.  \\This technique is useful for RV follow-up of transit surveys and for RV surveys of moderately rotating active stars looking for Jupiter mass planets.  It is also beneficial for looking at younger stars which tend to be faster rotating and therefore more active.  This could help improve our knowledge of planetary characteristics over time so we understand better how planets are formed and how they evolve.

\section*{Acknowledgments}
Victoria Moulds would like to thank the Northern Ireland Department for Employment and Learning for a PhD studentship. This work is based on data obtained from the ESO Science Archive Facility.  

\bibliographystyle{mn2e}
\bibliography{ref}

\begin{thebibliography}{}

\bibitem[\protect\citeauthoryear{Appourchaux, Michel, Auvergne, Baglin,
  Toutain, Baudi \& et al.}{Appourchaux et~al.}{2008}]{Appourchaux:2008di}
Appourchaux T.,  Michel E.,  Auvergne M.,  Baglin A.,  Toutain T.,  Baudi F.,
   et al. 2008, A\&A, 488, 705

\bibitem[\protect\citeauthoryear{Armitage \& Bonnell}{Armitage \&
  Bonnell}{2002}]{Armitage:2002di}
Armitage P.,  Bonnell I.,  2002, Monthly Notices of the Royal Astronomical
  Society, 330, L11

\bibitem[\protect\citeauthoryear{Barnes}{Barnes}{2004}]{Barnes:2004di}
Barnes J.~R.,  2004, Monthly Notices of the Royal Astronomical Society, 348,
  1295

\bibitem[\protect\citeauthoryear{Basri, Walkowicz, Batalha, Gilliland \& et
  al.}{Basri et~al.}{2010}]{Basri:2010di}
Basri G.,  Walkowicz L.~M.,  Batalha N.,  Gilliland R.,    et al. 2010, ApJ,
  713, L155

\bibitem[\protect\citeauthoryear{Benomar, Baudin, Campante, Chaplin \& et
  al.}{Benomar et~al.}{2009}]{Benomar:2009di}
Benomar O.,  Baudin F.,  Campante T.,  Chaplin W.,    et al. 2009, A\&A, 507,
  L13

\bibitem[\protect\citeauthoryear{Boisse, Bouchy, Hebrard, Bonfils, Santos \&
  Vauclair}{Boisse et~al.}{2011}]{Boisse:2011di}
Boisse I.,  Bouchy F.,  Hebrard G.,  Bonfils X.,  Santos N.,    Vauclair S.,
  2011, A\&A, 528

\bibitem[\protect\citeauthoryear{Boisse, Moutou, Vidal-Madjar, Bouchy \& et
  al.}{Boisse et~al.}{2009}]{Boisse:2009di}
Boisse I.,  Moutou C.,  Vidal-Madjar A.,  Bouchy F.,    et al. 2009, A\&A, 495,
  959

\bibitem[\protect\citeauthoryear{Borucki, Koch, Basri, Batalha \& et
  al.}{Borucki et~al.}{2011}]{Borucki:2011di}
Borucki W.,  Koch D.,  Basri G.,  Batalha N.,    et al. 2011, ApJ, 728, 117

\bibitem[\protect\citeauthoryear{Boss}{Boss}{1997}]{Boss:1997di}
Boss A.,  1997, Science, 276, 1836

\bibitem[\protect\citeauthoryear{Claret}{Claret}{2000}]{Claret:2000di}
Claret A.,  2000, A\&A, 363, 1081

\bibitem[\protect\citeauthoryear{Collier~Cameron, Donati \&
  Semel}{Collier~Cameron et~al.}{2002}]{Collier:2002di}
Collier~Cameron A.,  Donati J.-F.,    Semel M.,  2002, Monthly Notices of the
  Royal Astronomical Society, 330, 699

\bibitem[\protect\citeauthoryear{Desort, Lagrange, Udry \& Mayor}{Desort
  et~al.}{2007}]{Desort:2007di}
Desort M.,  Lagrange A.-M.,  Udry S.,    Mayor M.,  2007, A\&A, pp 983--993

\bibitem[\protect\citeauthoryear{Donati, Collier~Cameron, Hussain \&
  Semel}{Donati et~al.}{1999}]{Donati:1999di}
Donati J.,  Collier~Cameron A.,  Hussain G.,    Semel M.,  1999, Monthly
  Notices of the Royal Astronomical Society, 302, 437

\bibitem[\protect\citeauthoryear{Donati, Semel, Carter, Rees \&
  Collier~Cameron}{Donati et~al.}{1997}]{Donati:1997di}
Donati J.,  Semel M.,  Carter B.,  Rees D.,    Collier~Cameron A.,  1997,
  Monthly Notices of the Royal Astronomical Society, 291, 658

\bibitem[\protect\citeauthoryear{Faedi, Barros, Anderson, Brown,
  Collier~Cameron, Pollacco, Boisse \& et al.}{Faedi
  et~al.}{2011}]{Faedi:2011di}
Faedi F.,  Barros S.,  Anderson D.,  Brown D.,  Collier~Cameron A.,  Pollacco
  D.,  Boisse I.,    et al. 2011, A\&A, 531, A40

\bibitem[\protect\citeauthoryear{Figueira, Marmier, Bonfils, di Folco \& et
  al.}{Figueira et~al.}{2010}]{Figueira:2010di}
Figueira P.,  Marmier M.,  Bonfils X.,  di Folco E.,    et al. 2010, A\&A, 513,
  L8+

\bibitem[\protect\citeauthoryear{Gray}{Gray}{2008}]{Gray:2008di}
Gray D.,  2008, The Observation and Analysis of Stellar Photospheres.
Cambridge University Press

\bibitem[\protect\citeauthoryear{Grether \& Lineweaver}{Grether \&
  Lineweaver}{2006}]{Grether:2006di}
Grether D.,  Lineweaver C.,  2006, ApJ, 640, 1051

\bibitem[\protect\citeauthoryear{Hall \& Henry}{Hall \&
  Henry}{1994}]{Hall:1994di}
Hall D.,  Henry G.,  1994, International Amateur-Professional Photoelectric
  Photometry Communications, 55, 51

\bibitem[\protect\citeauthoryear{Hatzes, Dvorak, Wuchterl, Guterman \& et
  al.}{Hatzes et~al.}{2010}]{Hatzes:2010di}
Hatzes A.~P.,  Dvorak R.,  Wuchterl G.,  Guterman P.,    et al. 2010, A\&A,
  520, A93+

\bibitem[\protect\citeauthoryear{Hern{\'a}n-Obispo, G{\'a}lvez-Ortiz,
  Anglada-Escud{\'e}, Kane \& et al.}{Hern{\'a}n-Obispo
  et~al.}{2010}]{Hernan:2010di}
Hern{\'a}n-Obispo M.,  G{\'a}lvez-Ortiz M.,  Anglada-Escud{\'e} G.,  Kane S.,
   et al. 2010, A\&A, 512, A45+

\bibitem[\protect\citeauthoryear{Hu{\'e}lamo, Figueira, Bonfils, Santos \& et
  al.}{Hu{\'e}lamo et~al.}{2008}]{Huelamo:2008di}
Hu{\'e}lamo N.,  Figueira P.,  Bonfils X.,  Santos N.,    et al. 2008, A\&A,
  489, L9

\bibitem[\protect\citeauthoryear{Huerta, Johns-Krull, Prato, Hartigan \&
  Jaffe}{Huerta et~al.}{2008}]{Huerta:2008di}
Huerta M.,  Johns-Krull C.,  Prato L.,  Hartigan P.,    Jaffe D.,  2008, ApJ,
  678, 472

\bibitem[\protect\citeauthoryear{Jackson, Greenberg \& Barnes}{Jackson
  et~al.}{2008}]{Jackson:2008di}
Jackson B.,  Greenberg R.,    Barnes R.,  2008, in Sun Y.,  Ferraz-Mello S.,
  Zhou J.,  eds, IAU Symposium Vol.~249 of IAU Symposium, Tidal evolution of
  close-in extra-solar planets.
pp 187--196

\bibitem[\protect\citeauthoryear{Lagrange, Desort \& Meunier}{Lagrange
  et~al.}{2010}]{Lagrange:2010di}
Lagrange A.-M.,  Desort M.,    Meunier N.,  2010, Astronomy and Astrophysics,
  512, A38

\bibitem[\protect\citeauthoryear{Markwardt}{Markwardt}{2009}]{Markwardt:2009di}
Markwardt C.,  2009, in D.A.Bohlender D.Durand .~P.,  ed., Astronomical Data
  Analysis Software and Systems XVIII Vol.~411 of Astronomical Society of the
  Pacific Conference Series, Non-linear least-squares fitting in idl with
  mpfit.
pp 251--+

\bibitem[\protect\citeauthoryear{Mosser, Bouchy, Catala, Michel, Samadi \& et
  al.}{Mosser et~al.}{2005}]{Mosser:2005di}
Mosser B.,  Bouchy F.,  Catala C.,  Michel E.,  Samadi R.,    et al. 2005,
  A\&A, 431, L13

\bibitem[\protect\citeauthoryear{Paulson, Cochran \& Hatzes}{Paulson
  et~al.}{2004}]{Paulson:2004di}
Paulson D.,  Cochran W.,    Hatzes A.,  2004, ApJ, 127, 3579

\bibitem[\protect\citeauthoryear{Paulson \& Yelda}{Paulson \&
  Yelda}{2006}]{Paulson:2006di}
Paulson D.,  Yelda S.,  2006, The Publications of the Astronomical Society of
  the Pacific, 118, 706

\bibitem[\protect\citeauthoryear{Pollack, Hubickyj, Bodenheimer, Lissauer,
  Podolak \& Greenzweig}{Pollack et~al.}{1996}]{Pollack:1996di}
Pollack J.,  Hubickyj O.,  Bodenheimer P.,  Lissauer J.,  Podolak M.,
  Greenzweig Y.,  1996, Icarus, 124, 62

\bibitem[\protect\citeauthoryear{Queloz, Bouchy, Moutou, Hatzes \& et
  al.}{Queloz et~al.}{2009}]{Queloz:2009di}
Queloz D.,  Bouchy F.,  Moutou C.,  Hatzes A.,    et al. 2009, A\&A, 506, 303

\bibitem[\protect\citeauthoryear{Queloz, Henry, Sivan, Baliunas \& et
  al.}{Queloz et~al.}{2001}]{Queloz:2001di}
Queloz D.,  Henry G.,  Sivan J.,  Baliunas S.,    et al. 2001, A\&A, 379, 279

\bibitem[\protect\citeauthoryear{Ricker, Latham, Vanderspek, Ennico, Bakos,
  Brown, Burgasser, Charbonneau, Clampin \& et al.}{Ricker
  et~al.}{2010}]{Ricker:2010di}
Ricker G.,  Latham D.,  Vanderspek R.,  Ennico K.,  Bakos G.,  Brown T.,
  Burgasser A.,  Charbonneau D.,  Clampin M.,    et al. 2010, in American
  Astronomical Society Meeting Abstracts 215 Vol.~42 of Bulletin of the
  American Astronomical Society, Transiting exoplanet survey satellite (tess).
p. 450.06

\bibitem[\protect\citeauthoryear{Saar \& Donahue}{Saar \&
  Donahue}{1997}]{Saar:1997di}
Saar S.,  Donahue R.,  1997, ApJ, 485, 319

\bibitem[\protect\citeauthoryear{Setiawan, Henning, Launhardt, M{\"u}ller \& et
  al.}{Setiawan et~al.}{2008}]{Setiawan:2008di}
Setiawan J.,  Henning T.,  Launhardt R.,  M{\"u}ller A.,    et al. 2008,
  Nature, 451, 38

\bibitem[\protect\citeauthoryear{Strassmeier, Lupinek, Dempsey \&
  Rice}{Strassmeier et~al.}{1999}]{StrassmeierA:1999di}
Strassmeier K.,  Lupinek S.,  Dempsey R.,    Rice J.,  1999, A\&A, 347, 212

\bibitem[\protect\citeauthoryear{Valenti \& Fischer}{Valenti \&
  Fischer}{2005}]{Valenti:2005di}
Valenti J.,  Fischer D.,  2005, ApJ, 159, 141

\bibitem[\protect\citeauthoryear{Watson, Dhillon \& Shahbaz}{Watson
  et~al.}{2006}]{Watson:2006di}
Watson C.,  Dhillon V.,    Shahbaz T.,  2006, Monthly Notices of the Royal
  Astronomical Society, 368, 637

\end{thebibliography}

\label{lastpage}

\end{document}